\documentclass[fleqn,usenatbib]{mnras}

\usepackage[T1]{fontenc}

\DeclareRobustCommand{\VAN}[3]{#2}
\let\VANthebibliography\thebibliography
\def\thebibliography{\DeclareRobustCommand{\VAN}[3]{##3}\VANthebibliography}

\usepackage{graphicx}	
\usepackage{amsmath}	
\usepackage{amssymb}	
\usepackage{subcaption}
\usepackage[dvipsnames]{xcolor}
\usepackage{tikz}       
\usepackage{newtxtext,newtxmath}

\newcommand{\rev}{}
\newcommand{\revb}{}

\title[Cross-Identification Using RL-Xid]{The Application of Ridgelines in Extended Radio Source Cross-Identification}

\author[B. Barkus et al.]{
B. Barkus,$^{1}$\thanks{E-mail: bonny.barkus@open.ac.uk}
J. H. Croston,$^{1}$
J. Piotrowska,$^{2,3}$
B. Mingo,$^{1}$
P. N. Best,$^{4}$
\newauthor
M. J. Hardcastle,$^{5}$
R. I. J. Mostert,$^{6,7}$
H. J. A.  R\"ottgering,$^{6}$
J. Sabater,$^{4}$
\newauthor
B. Webster$^{1}$
W. L. Williams,$^{6}$
\\
$^{1}$School of Physical Sciences, The Open University, Walton Hall, Milton Keynes, MK7 6AA, UK\\
$^{2}$Cavendish Laboratory, Astrophysics Groups, University of Cambridge, 9 JJ Thomson Avenue, Cambridge, CB3 0HE, UK\\
$^{3}$Kavli Insitute for Cosmology, Madlingley Road, CB3 0HA, Cambridge, UK\\
$^{4}$Institute for Astronomy, University of Edinburgh, Royal Observatory, Blackford Hill, Edinburgh, EH9 3HJ, UK\\
$^{5}$Centre for Astrophysics Research, University of Hertfordshire, College Lane, Hatfield, AL10 9AB, UK\\
$^{6}$Leiden Observatory, Leiden University, PO Box 9513, NL-2300 RA Leiden, The Netherlands\\
$^{7}$ASTRON, the Netherlands Insitute for Radio Astronomy, Postbus 2, 7990 AA, Dwingeloo, The Netherlands\\
}

\date{Accepted XXX. Received YYY; in original form ZZZ}

\pubyear{2020}

\begin{document}
\label{firstpage}
\pagerange{\pageref{firstpage}--\pageref{lastpage}}
\maketitle

\begin{abstract}
Extended radio sources are an important minority population in modern deep radio surveys, because they enable detailed investigation of the physics governing radio-emitting regions such as active galaxies and their environments. Cross-identification of radio sources with optical host galaxies is challenging for this extended population, due to their morphological complexity and multiple potential counterparts. In the first data release of the Low-frequency array (LOFAR) Two-metre Sky Survey (LoTSS DR1) the automated likelihood ratio for compact sources was supplemented by a citizen science visual identification process for extended sources. In this paper we present a novel method for automating the host identification of extended sources \revb{by using} ridgelines, which trace the assumed direction of fluid-flow through the points of highest flux \rev{density}.  Applying a new code, RL-Xid, to LoTSS DR1, we demonstrate that ridgelines are versatile; by providing information about spatial structure and brightness distributions, they can be used both for optical host identification and morphological studies in radio surveys. \revb{RL-Xid draws ridgelines for 85 per cent of sources brighter than 10 mJy and larger than 15 arcsec, with an improved performance of 96 per cent for the subset >30 mJy and >60 arcsec.} Using a sample of sources with known hosts from LoTSS DR1, we demonstrate that RL-Xid successfully identifies the host for 98 per cent of the sources with successfully drawn ridgelines, and performs at a comparable level to visual identification via citizen science. We also demonstrate that ridgeline brightness profiles provide a \rev{promising} automated technique for morphological classification.

\end{abstract}

\begin{keywords}
galaxies: active -- galaxies: jets -- methods: statistical -- radio continuum: galaxies -- software: data analysis -- software: development
\end{keywords}



\section{Introduction}

In order to better understand the role of active galactic nuclei (AGN) jets in galactic feedback processes one requires an in-depth knowledge of jet interactions with their surrounding medium \citep[e.g.][]{Hardcastle2019}.  Such jets are found in radio galaxies, where they demonstrate their presence through the synchrotron radiation emitted by charged particles carried out to large distances.  AGN jets can reach megaparsecs in size and are extremely powerful in terms of kinetic luminosity and radiative signatures, emitting over the entire electromagnetic spectrum \citep{Worrall2009, Hardcastle2016}.  Observations of jet interactions with their environment can reveal properties such as velocity, energy density and magnetic field content, while measurements of the jet parameter can be used to make inferences about the environment itself \citep[e.g.][]{Mao1987}.  Properly characterising physical properties such as age and size for the extended sources in a radio galaxy population allows models of radio galaxy impact to be fully tested \citep[e.g.][]{Hardcastle2019}.

Surveys such as EMU (Evolutionary Map of the Universe) with ASKAP (Australian Square Kilometre Array Pathfinder), \citep{NorrisA2011}, MIGHTEE (The MeerKAT International GHz Tiered Extragalactic Exploration) Survey with MeerKAT \citep{Jarvis2017} and LoTSS (The LOFAR Two-metre Sky Survey) with LOFAR \citep{Shimwell2019} as SKA (Square Kilometre Array) precursors are expected to observe millions of sources each when completed.  The science goals of radio continuum surveys are to understand many aspects of the universe including the evolution of active galactic nuclei \citep{Smolcic2016}, star formation \citep{Beswick2016, Shimwell2017}, and galaxy clusters \citep{Shimwell2017, Hardcastle2019}.  For all of the scientific aims of these surveys, redshifts are needed and must be obtained by finding the counterparts to radio sources in optical/infrared (IR) surveys.

Cross-identification is the matching of radio sources with optical/IR sources from other surveys, and it is an essential step towards obtaining the redshifts that are needed to determine physical quantities such as luminosity and source size.  This essential task is made difficult because of the extended nature of some radio sources \citep{Williams2019, Kondapally2021}.  A further challenge for many radio surveys in obtaining redshifts is the lack of high-quality photometric data to determine good photometric redshifts.  AGN are harder to obtain reliable photometric redshifts for and a method which works for one astronomic population (e.g. AGN) will not be as successful on another \citep{Duncan2019, Salvato2019}.  However, with wide area data such as PanSTARRS and AllWISE, photometric redshifts can be determined provided the radio source host galaxy can be identified \citep{Duncan2019}.  Upcoming radio surveys present even greater cross-identification challenges due to the increased volume of data. 

Identification of possible host candidates, \rev{when cross comparing catalogues with a range of resolutions,  has often been carried out} using the Nearest Neighbour technique \citep{Kimball2008}.  Although successful for some earlier surveys, this is not practical for cross-identification over multiple wavelengths as this method can produce multiple possible matches to extended sources where the nearest neighbour ID can be wrong \citep{Kondapally2021}.  The likelihood ratio method \rev{\citep{Richter1975}} helps by statistically determining when to accept a possible counterpart depending on the properties of the radio source and the potential optical/IR host.  With the larger catalogues being produced by new and upcoming surveys, the increased volume of data will lead to a higher number of extended sources (or potentially multi-component sources) observed, and multiple matches to be interpreted.  

Automated methods are needed to deal with the great volume of data and to improve the characterisation of extended sources.  Methods of cross-identification include variations on the likelihood ratio method \citep[e.g.][]{Richter1975};  this class of methods uses statistical determination of the likelihood of an optical/IR source being a matching counterpart to a radio source, based on parameters including separation, colour and magnitude of the sources, along with their associated errors. There also exist techniques using Poisson Probability \cite{Downes1985}, and Bayesian methods of hypothesis testing \cite{Fan2015} for component matching of radio sources with realistic morphology.  The success of the likelihood ratio method can be limited by sources which are too large and complex \citep[see][]{Pineau2016}.  In some circumstances the source finder is unable to correctly group separate components of large sources, or incorrectly groups multiple sources together, and the likelihood ratio is unable to correct for this \citep{Williams2019}.  In these instances when automated methods cannot be used, other visual classifications have been developed, such as Radio Galaxy Zoo \citep{Banfield2015}, where citizen science is used to identify host galaxies.  This can be time consuming, as multiple classifications are needed per radio source.

In this paper we consider a novel idea for automated host identification, and potentially morphological classification, of extended radio sources: the use of spatial information in the form of ridgelines.  \rev{From the late 1970's the concept of tracing the direction of fluid flow along a jet using the radio brightness has been used, especially in the context of dynamical modelling (see \cite{Blandford1978, Icke1981, Gower1982, Condon1984, Hunstead}).  More sophisticated ridgeline analysis and algorithmic approaches, such as the one presented in this paper, are largely associated with VLBI studies (see \cite{Britzen2010, Karouzos2012, Perucho2012, Vega-Garcia2019}).  Similar techniques have been used to study other types of source, such as the x-ray binary jet SS433 \cite[e.g.][]{Blundell2004}.  Total intensity ridgelines based on the method of \cite{Pushkarev2017}, where the integrated intensity is equal on both sides of the ridgeline, have been used by multiple authors to investigate properties of the ridgelines themselves, such as their spectral index and the electric vector position angle (see \cite{Li2018, Pushkarev2019, Kravchenko2020, Lico2020}). }

Morphological classification of extended radio sources is another process which has recently been automated, for example by LoMorph \citep{Mingo2019}\footnote{\url{https://github.com/bmingo/LoMorph}} and plays an important role in the understanding of feedback processes.  Traditionally, radio-loud AGN are divided into two Fanaroff and Riley (FR) morphologies; FRI and FRII.  These are defined based on the ratio of the distance between the region of highest brightness and the core on each side, and the full length of the corresponding side.  If the ratio < 0.5, where the peak of the brightness is near the core, it is an FRI and if the brightest peak is near the edge of the lobes, giving a ratio > 0.5, it is an FRII \citep{Fanaroff1974}.  Surface brightness profiles along ridgelines therefore have the potential to be used as a means to classify extended radio sources.

In this work we make use of surface brightness information via ridgelines to improve the automated likelihood ratio method for host identification, and reduce the number of sources going to human classification such as Radio Galaxy Zoo.  In this way we both speed up the identification of an optical/IR counterpart and lower the chances of human error associated with visual inspection.  We also extend our investigation to automated morphological classification, further demonstrating the versatility of ridgelines in radio astronomy.  We introduce the radio and optical/IR catalogues in Section \ref{sec:Data}, followed by ridgeline code introduction and examples given in Section \ref{sec:RLC} \rev{(with more details given in the Appendix \ref{SubSec:Results})
}. We then use ridgelines to define and generate Surface Brightness Profiles in Section \ref{sec:SB}. We follow this by applying both ridgelines and the optical host properties, magnitude and colour, in Section \ref{sec:LR}.  We summarise our results in Section \ref{sec:Sum}.

\section{Data}
\label{sec:Data}

We wish to investigate whether a ridgeline-based host identification method can obtain results comparable to those from visual analysis of extended sources.  The LoTSS DR1 value-added catalogue \citep{Williams2019} provides a large set of extended radio sources that have been classified using the LOFAR Galaxy Zoo (LGZ)\footnote{\url{https://www.zooniverse.org/projects/chrismrp/radio-galaxy-zoo-lofar}} citizen science approach, built on the Zooniverse platform.  Although a citizen science approach was used for DR1, the classifications were performed by experts in the field which is likely to have \rev{improved the reliability of} the results.  In this work, we therefore adopt the DR1 LGZ identifications as the truth set used for comparing our results.  \rev{We have further tested the reliability of the LGZ truth set by examining the subset of our sample of large, bright sources for which a compact FIRST source consistent with a radio core is present (62 per cent of our sample), finding that for 97 per cent of sources with a bright, compact FIRST source in a plausible host location, the LGZ host ID coincides with the FIRST core.  This is not surprising, as the FIRST contours were available as part of the LGZ visual process.}  \rev{We are therefore confident that it forms a useful} truth set that represents the state of the art in host identification for large extended source samples, and so provides the best testing ground for our method, although we note that the LGZ catalogue is not perfect, with a small number of source matches resulting in incorrect IDs.  As discussed in Section \ref{sec:Sum}, optimising the method for rejection of sources with no valid host will be the subject of future work with the upcoming second data release of LoTSS.

The data is taken from LoTSS DR1 \citep{Shimwell2019, Williams2019}, a 120 – 168 MHz continuum survey which covers 424 deg$^2$ of the northern hemisphere at a resolution of 6 arcsec (giving 1 pixel $\approx$ 1.5 arcsec). LoTSS DR1 contains 318,520 sources where 73 per cent have optical counterparts \citep{Williams2019}.  We selected our final sample from the 23,344 sources defined as radio loud AGN by \cite{Hardcastle2019} and compared against the combined optical/IR Pan-STARRS and AllWISE catalogue used by \cite{Williams2019}.  We chose to focus on AGN during the method development stages to avoid contamination from large star-forming galaxies.   We note, however, that initial testing indicates that while the method is motivated by jet morphologies, it also performs successfully on extended star-forming galaxies, because their emission is typically symmetric about the major axis with a centrally located peak.  The \cite{Williams2019} dataset includes fully associated components and so it is important to remember that the problem of associating components is beyond the scope of this paper.  It is assumed this will be done by other means prior to the application of our methods, although there is the potential of extending the use of ridgelines to this problem as well.

\begin{figure*}
    \centering
    \includegraphics[width = 0.85\linewidth]{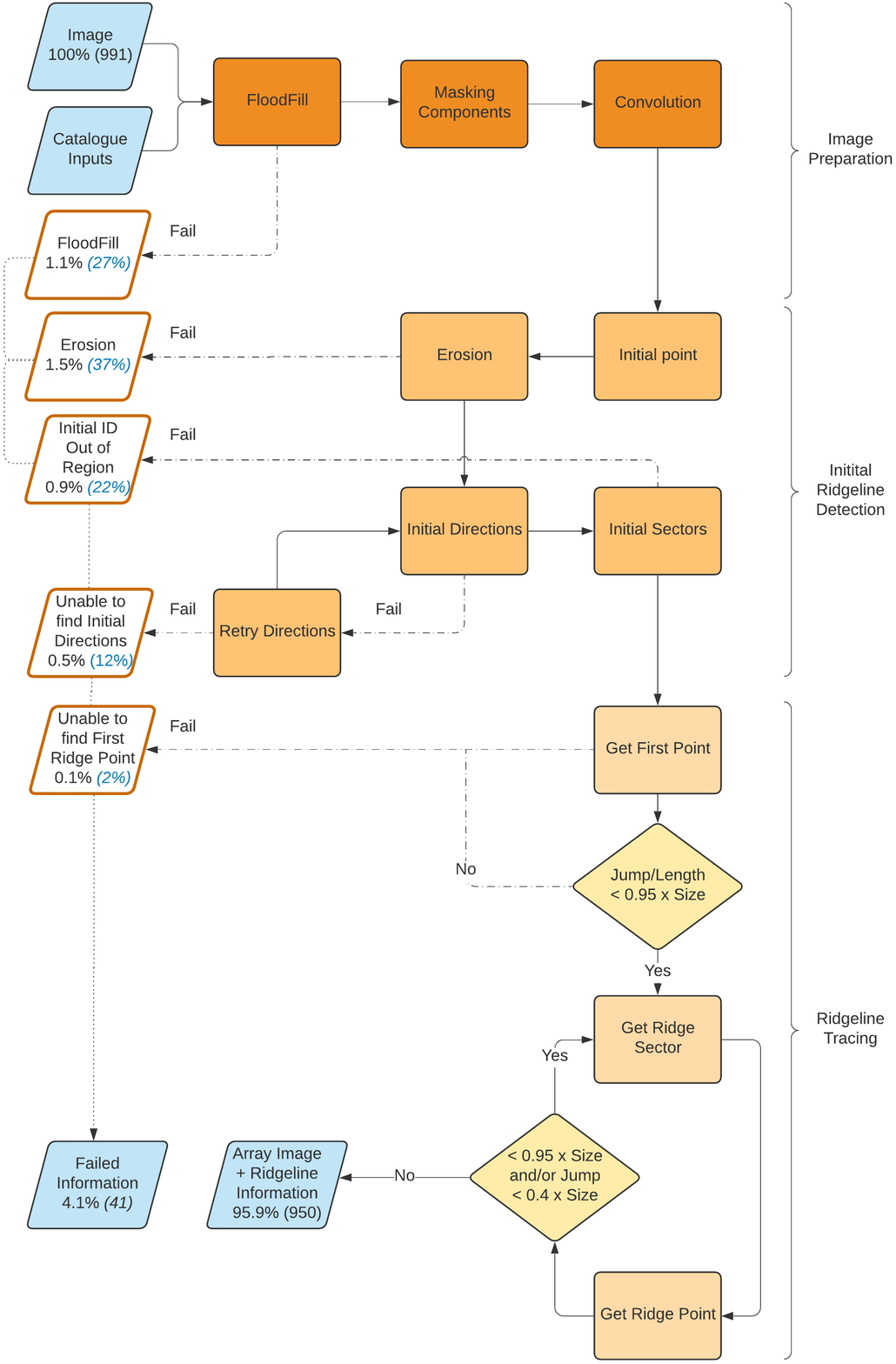}
    \caption{The workflow of RL-Xid. There are three main stages in the code: Image Preparation, Initial Ridgeline Detection and Ridgeline Tracing.  The various routes by which the code can produce an error and pass the source out as a \revb{\textit{Failed}} are shown with the dashed lines.  The different processes carried out by the code are shown in the rectangular boxes with the decision restrictions in the diamond boxes. The full sample size is given in the Image box and the number of sources passed out as each type of \revb{\textit{Failed}} is given, as a percentage of the full sample (black text) and as a percentage of the total number of \revb{\textit{Failed}} (blue text).  The final number of \textit{\revb{Completed}} and \textit{\revb{Failed}} sources is given as a percentage of the full sample. A detailed description of the process is given in Section \ref{SubSec:RLXM}}
    \label{fig:flow}
\end{figure*}

The shape and sizes of the sources were obtained and catalogued using the Python source detection code, PyBDSF \citep{Mohan2015, Shimwell2019, Williams2019}.  For LoTSS DR1 \cite{Williams2019} used a decision tree to select which sources were to have their hosts determined automatically and which went to a visual classification.  The large and bright radio sources, most likely to require visual classification, were defined to have a major axis $> 15$ arcsec and flux density $> 10$ mJy.  Those sources smaller than 15 arcsec with nearby companions were considered as possible unassociated components of complex sources, and were inspected again before being sent to LGZ for visual classification if needed.

We carried out initial development of the method, using a dataset containing 991 of the largest and brightest sources (with total flux \rev{density} $> 30$ mJy and a size $> 60$ arcsec, described in Sections \ref{sec:RLC} through to \ref{SubSec:SP}).  \revb{Source size is a necessary input parameter for our method (see Section \ref{SubSec:RLXM}); however, the \cite{Williams2019} catalogued sizes are only rough approximations for complex extended sources in which multiple PyBDSF components have been associated.  We therefore, instead, make use of the source sizes determined by \cite{Mingo2019} using the LoMorph code.  The \cite{Mingo2019} sizes are calculated from an image thresholded at the higher of 4 times the RMS or $1/50$ of the peak brightness by determining the greatest distance between two non-zero pixels assumed to be part of the source.}  We later explore the performance of our method over a wider range in source parameters, as described fully in Section \ref{subsec:OutLR}.

In order to ensure a fair comparison between the performance of our method and the identification methods of \cite{Williams2019} and LGZ, we use the same parent catalogue of potential host galaxies.  The optical/IR catalogue includes the sources from Pan-STARRS 3$\pi$ survey \citep{Chambers2016} and the AllWISE catalogue \citep{Wright2010} in the same sky area as LoTSS DR1.  In total there are over 26.5 million sources which are detected in either Pan-STARRS, AllWISE or both \citep{Williams2019}.  The potential hosts in the radio sample are allowed a maximum optical error for both RA and DEC of $< 0.3$ arcsec.  This selection excludes a small subset (105,407, 0.395 per cent) of potential hosts with poorly constrained positions (RA and DEC errors over $1$ arcsec) from the optical/IR catalogue which could affect the cross-matching probability.  Throughout this work magnitudes are in the AB system \citep{Oke1983}.

\section{Method for Ridgeline Characterization}
\label{sec:RLC}
Using a similar approach to \cite{Pushkarev2017}, we wrote a new code (RL-Xid) 
to generate total radio emission intensity ridgelines and investigate their application towards optical host identification and morphological classification of radio sources.  A ridgeline is defined as a piecewise linear pathway of connected points of highest intensity, separated by a full width of a telescope beam.

In the case of AGN jets, a ridgeline is intended to trace the direction of fluid flow. Under the assumption that jets are generated by the supermassive black holes residing at galactic centres, \rev{and that the radio emission is typically laterally symmetric about the direction of the jet flow}, each ridgeline should pass through \rev{(or near to)} the centre of its optical/IR host.

\subsection{Methods}
\label{SubSec:RLXM}
\begin{figure}
    \begin{subfigure}{0.9\columnwidth}
        \centering
        \includegraphics[width=0.8\columnwidth]{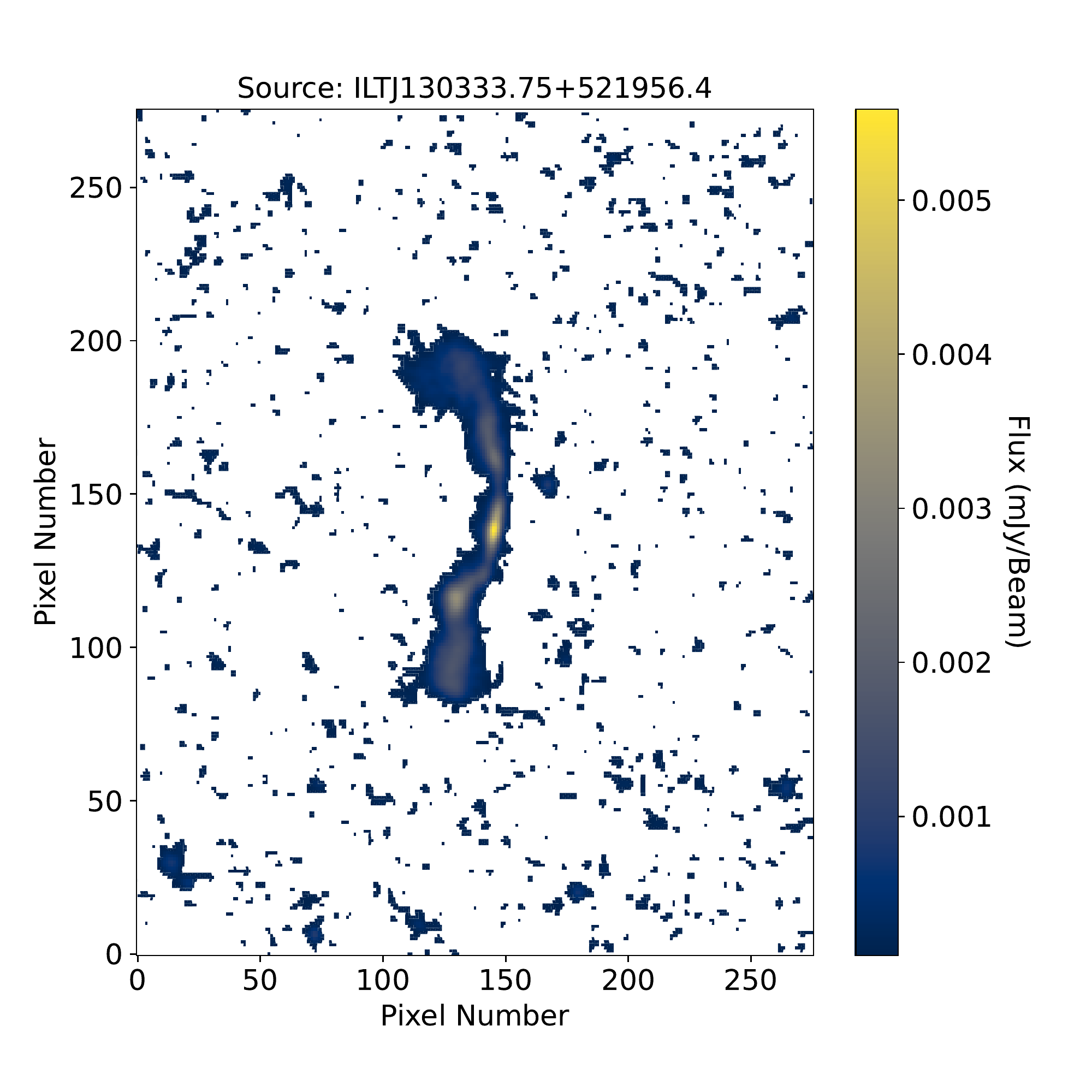}
        \caption{\label{fig:ExSource}}
    \end{subfigure}
    \begin{subfigure}{0.9\columnwidth}
        \centering
        \includegraphics[width=0.8\columnwidth]{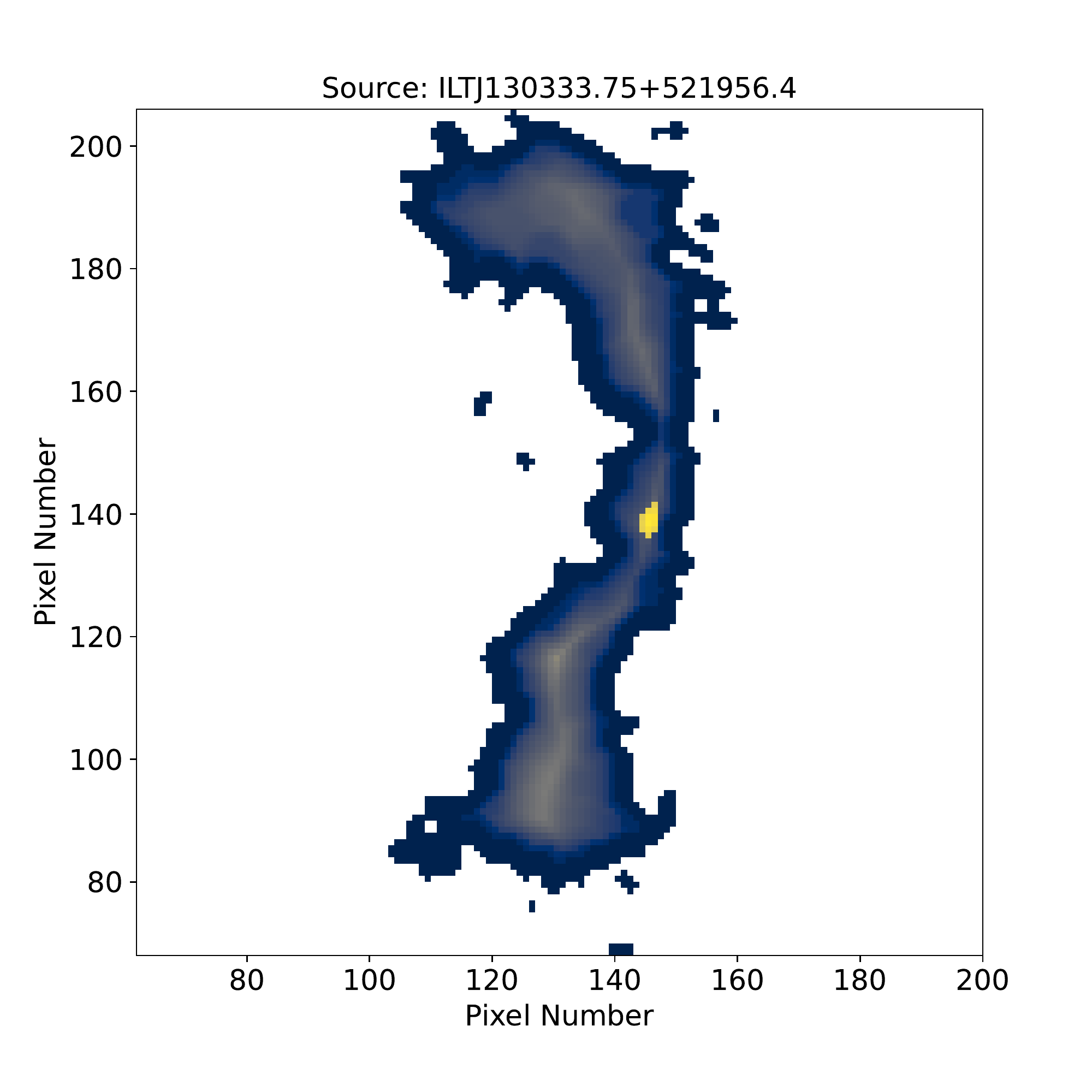}
        \caption{\label{fig:ExEroded}}
    \end{subfigure}
    \begin{subfigure}{0.9\columnwidth}
        \centering
        \includegraphics[width=0.8\columnwidth]{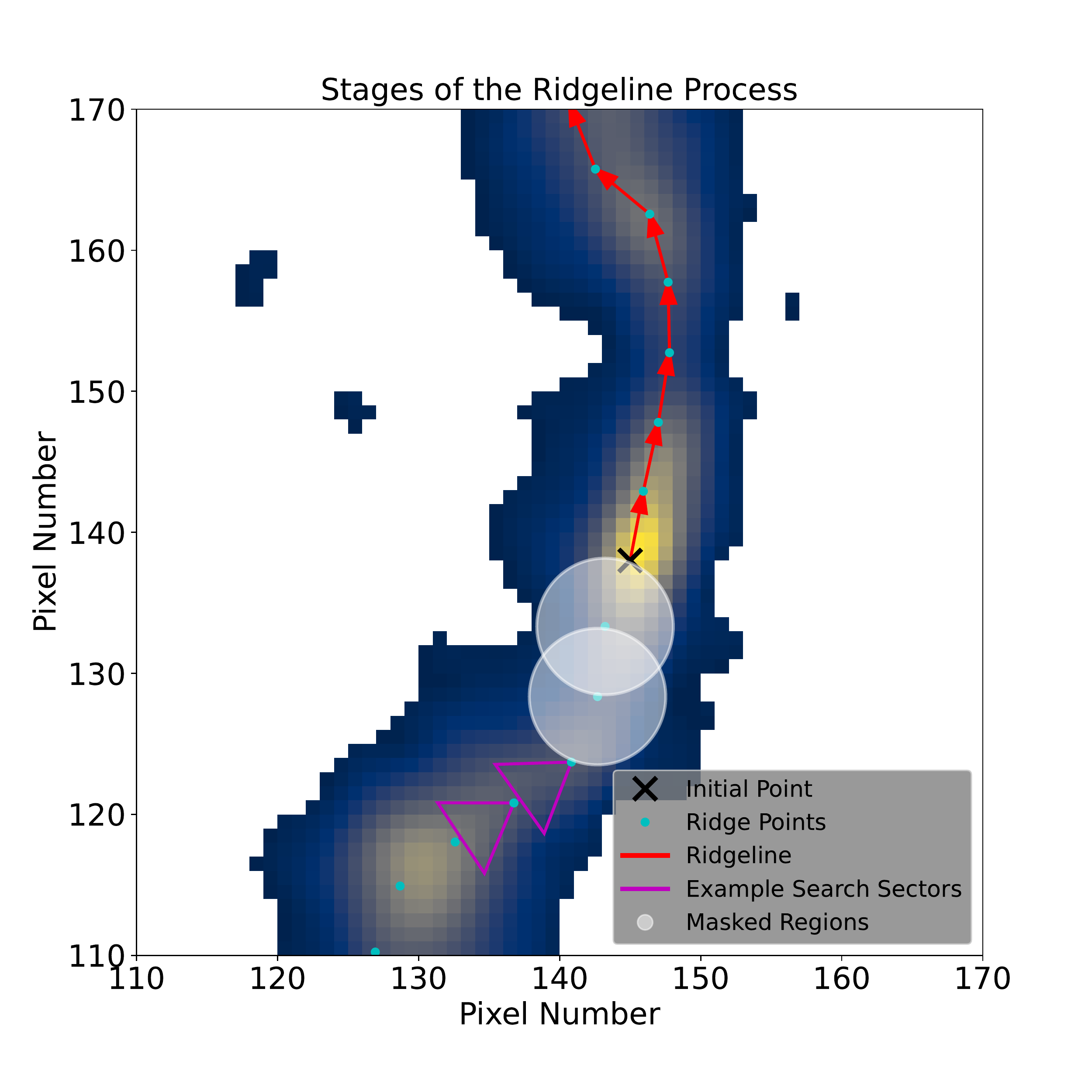}
        \caption{\label{fig:ExStages}}
    \end{subfigure}
    \caption{Three images showing the process of generating a ridgeline \revb{(1 pixel $\approx$ 1.5 arcsec)}.  (a): the \rev{source} cut-out before the ridgeline process takes place; (b): a close up of the source after it has been eroded, \rev{showing enhancement of the brightest regions}; (c): \rev{a zoomed-in image to illustrate the} different stages of the ridgeline process (see text for full description). The black cross is the initial starting point, the magenta triangles represent the search \revb{sectors}, the cyan dots represent the located ridge points, the white circles represent the masked out areas around the previous points and the red arrows show how the points are joined together to form the final ridgeline.}
    \label{fig:Examples}
\end{figure}

RL-Xid\footnote{\url{https://github.com/BonnyBlu/RL-Xid/tree/main/LOFAR/DR1}} is written in Python using standard packages as well as Sci-Kit Image\footnote{\url{https://scikit-image.org/}} \citep{VanDerWalt2014}, Astropy\footnote{\url{https://www.astropy.org/}} \citep{Collaboration2013, Price-Whelan2018}, and PyRegion\footnote{\url{https://pyregion.readthedocs.io/en/latest/}}.  Figure \ref{fig:flow} shows the workflow of the code involving three main parts, each of which are described below in detail. \rev{RL-Xid acts on a set of input image arrays, to which a threshold has been applied to mask extraneous emission. The threshold is set at the higher of (i) 4 times the locally measured RMS noise, or (ii) $1/50^{th}$ of the source’s peak flux density. The latter dynamic range criterion was found to provide an optimal threshold for image analysis of bright sources in this dataset by \cite{Mingo2019}.}  There are two outputs for each input source, text files tracing the ridgeline and an image of the source with the ridgeline traced on top of it. If a source has a successfully drawn ridgeline \rev{it is defined as a \textit{\revb{Completed}} source, and} the code generates two joint text files, each containing information on the location of the ridge points, the angular difference in radians between two consecutive points and the length of the ridgeline in pixels.  The code also generates an image of the source after the image preparation steps described in the following Section, with the ridge points overlaid and joined to form the ridgeline (see Figure \ref{fig:Examples}). If the code fails at any point during the process the source is deemed a \textit{\revb{Failed}} \rev{source} and the processed image is stored and a file containing a list of all the sources and type of error is produced.  The key steps in the process are shown in Figure \ref{fig:flow} and Figure \ref{fig:Examples}.

\subsubsection{Image Preparation}
The image is first prepared to remove any non-associated sources and extraneous emission.  This confines the ridgeline process to \rev{emission} contained within the source of interest.  The component masking and \texttt{FloodFill} codes from LoMorph \citep{Mingo2019} are applied in order to \rev{mask nearby non-associated PyBDSF components, and remove extraneous emission not related to associated components from the cut-out}.  \revb{The component masking and flood-fill stage enables the masking of all emission not associated with the source, and is explained in full detail in Section 2.2 of \cite{Mingo2019}.  In summary, a mask is made that contains all pixels within the \cite{Williams2019} catalogued Gaussian components associated with the source, and excludes any nearby components catalogued as not associated; the Python skimage.measure label\footnote{\url{https://scikit-image.org/docs/dev/api/skimage.measure.html\#skimage.measure.label}} routine is then used to identify connected islands of emission, which are used to extend the source's outer boundary beyond the Gaussian components to include any connected emission.  In this way a source mask is generated that excludes} background noise, artefacts and faint uncatalogued sources, as well as nearby bright sources.  This careful masking of unassociated emission from around the source is essential to ensure the ridgeline traces only associated emission and does not extend beyond the boundary of the source.  Finally, the image is smoothed through a convolution using a cross shaped, centre weighted kernel.  The convolution is performed to help remove any localized \rev{edge effects due to the component masking}.   \rev{\texttt{Erosion} is an skimage.morphology function\footnote{\url{https://scikit-image.org/docs/dev/api/skimage.morphology.html\#skimage\\.morphology.erosion}} which is performed, using an octagonal kernel,  to emphasise the brightest regions in the image above a given threshold.}  \revb{The octagonal kernel represents the LOFAR beam shape, and as the erosion function minimises the pixel values over this area (centred on the centre pixel), this highlights the brighter areas whilst enlarging the fainter ones.}  The erosion and convolution kernels were carefully tested to make sure that the optimal size and weight were used to balance the amount of image eroded and noise remaining.  A kernel size of just below beam width was found as optimal; see Figure \ref{fig:flow} and Figure \ref{fig:Examples}.

\begin{figure*}
    \begin{subfigure}{0.5\linewidth}
        \centering
        \includegraphics[width=\linewidth]{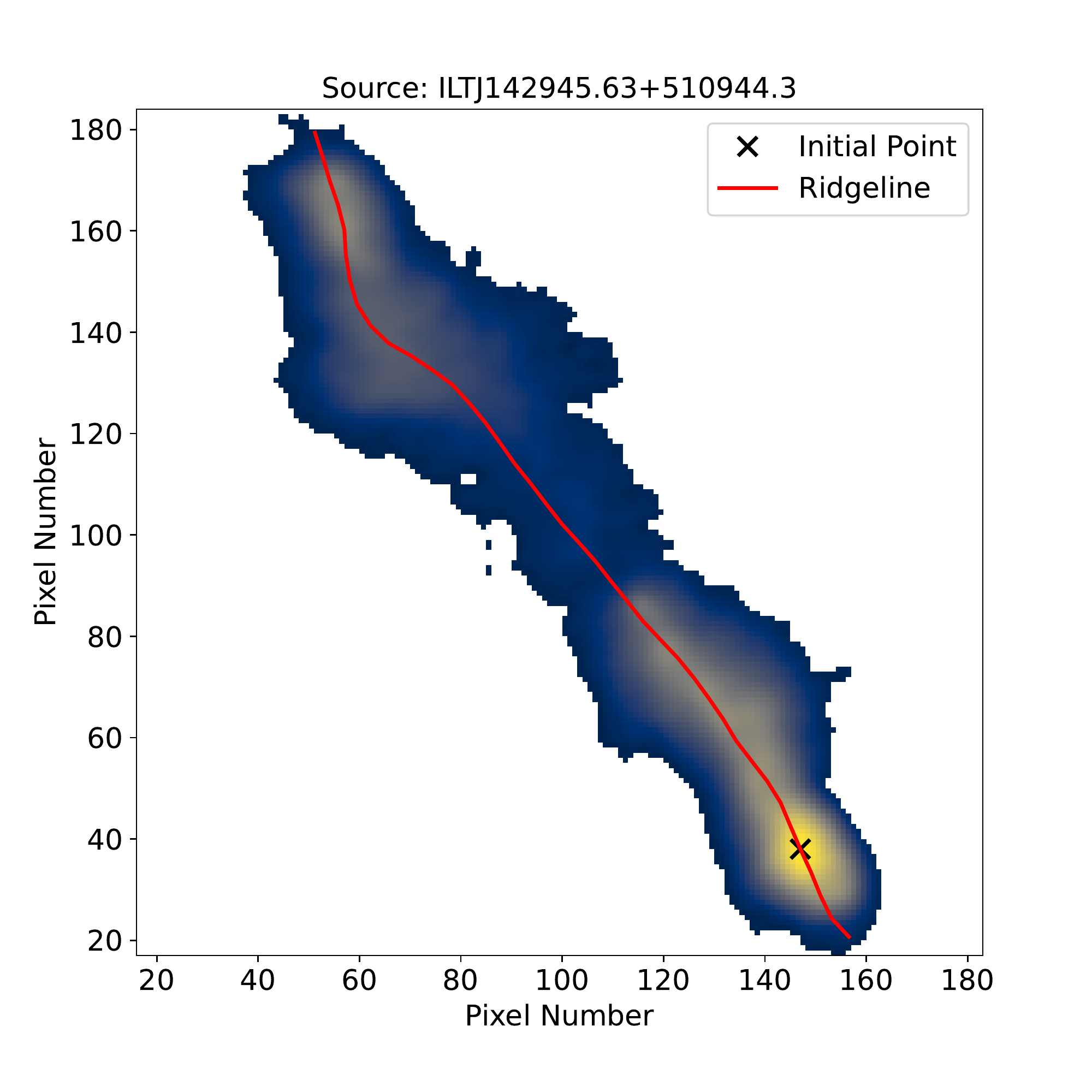}
        \caption{\label{fig:SucRL}}
    \end{subfigure}
    \begin{subfigure}{0.5\linewidth}
        \centering
        \includegraphics[width=\linewidth]{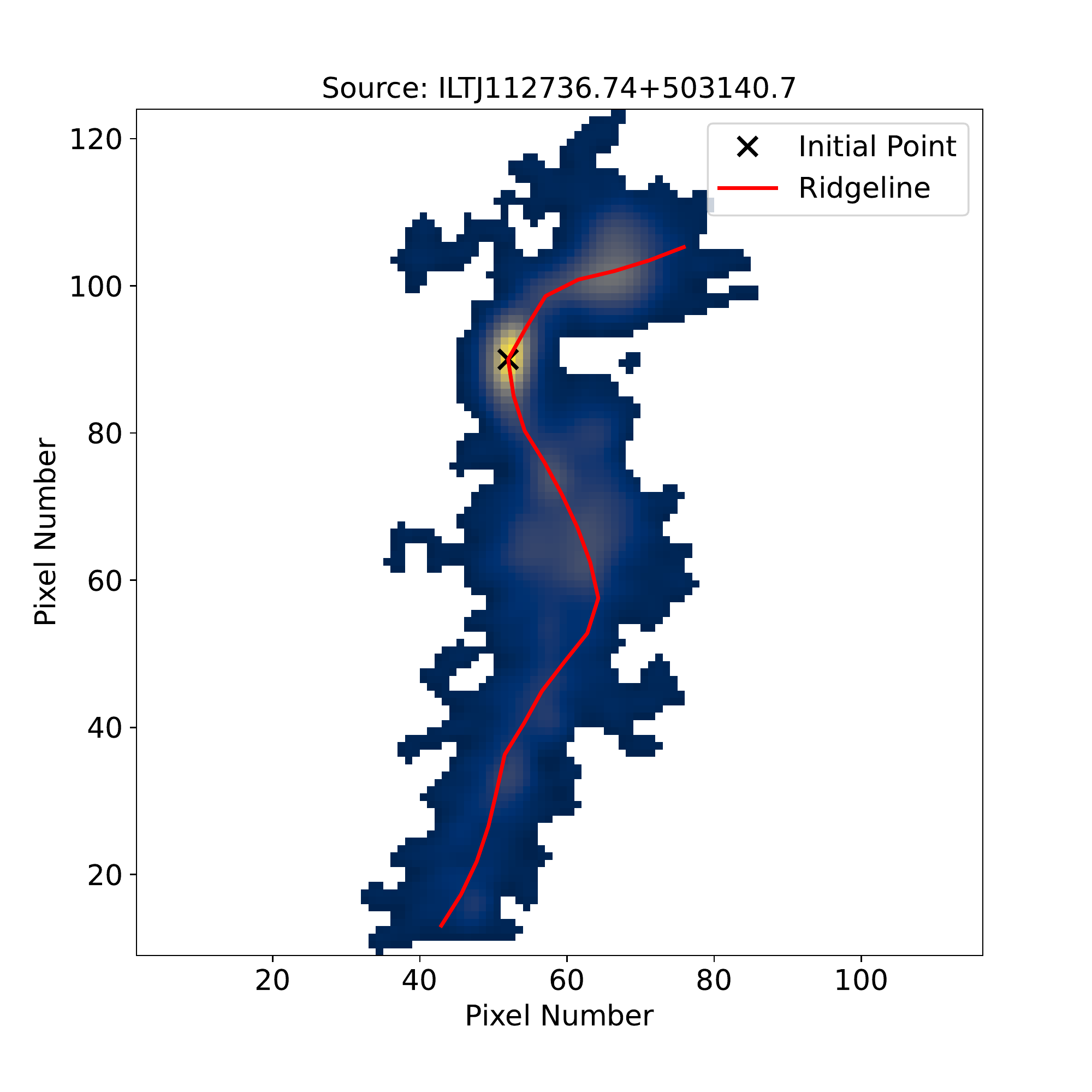}
        \caption{\label{fig:SucBent}}
    \end{subfigure}
    \begin{subfigure}{0.5\linewidth}
        \centering
        \includegraphics[width=\linewidth]{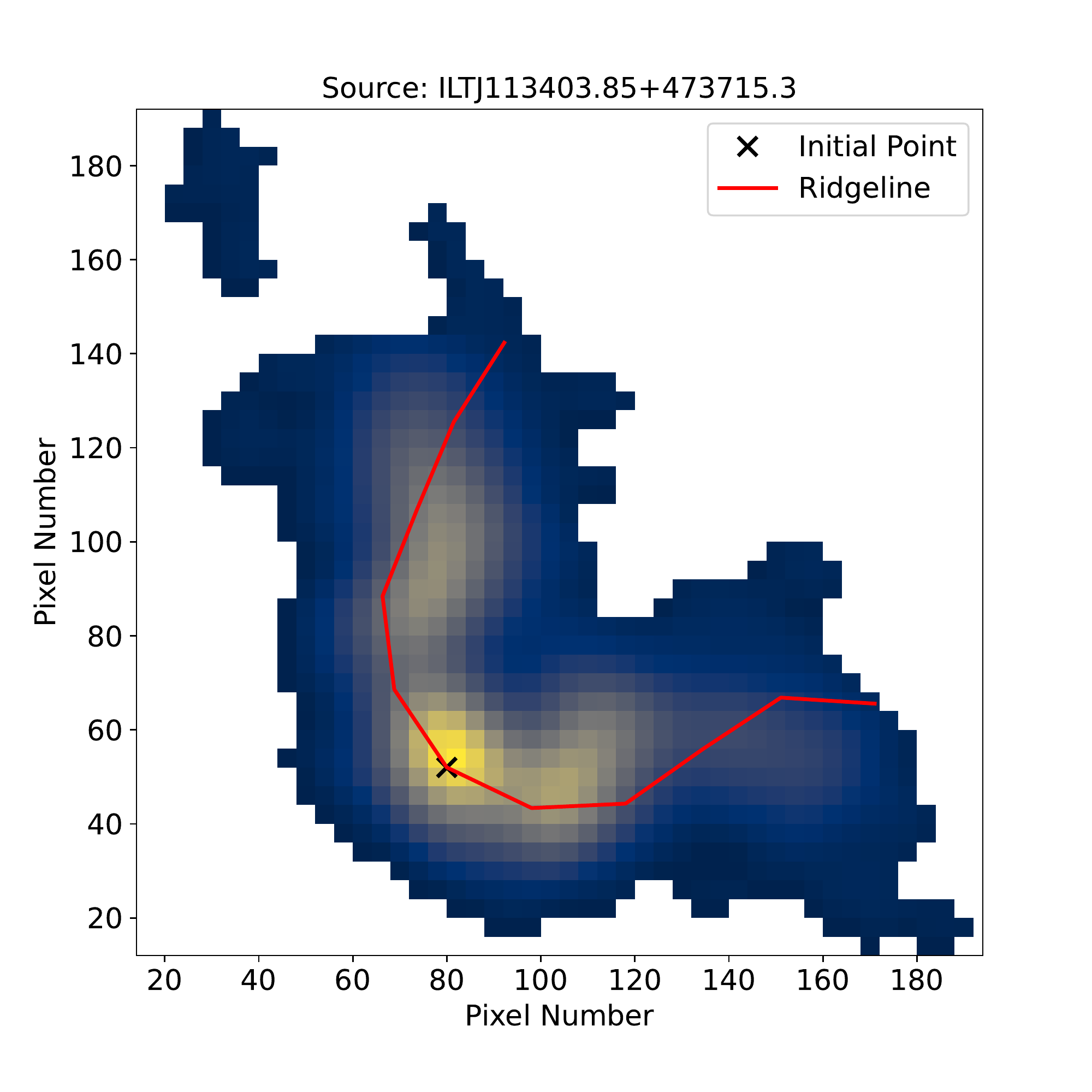}
        \caption{\label{fig:SucWAT}}
    \end{subfigure}
    \begin{subfigure}{0.5\linewidth}
        \centering
        \includegraphics[width=\linewidth]{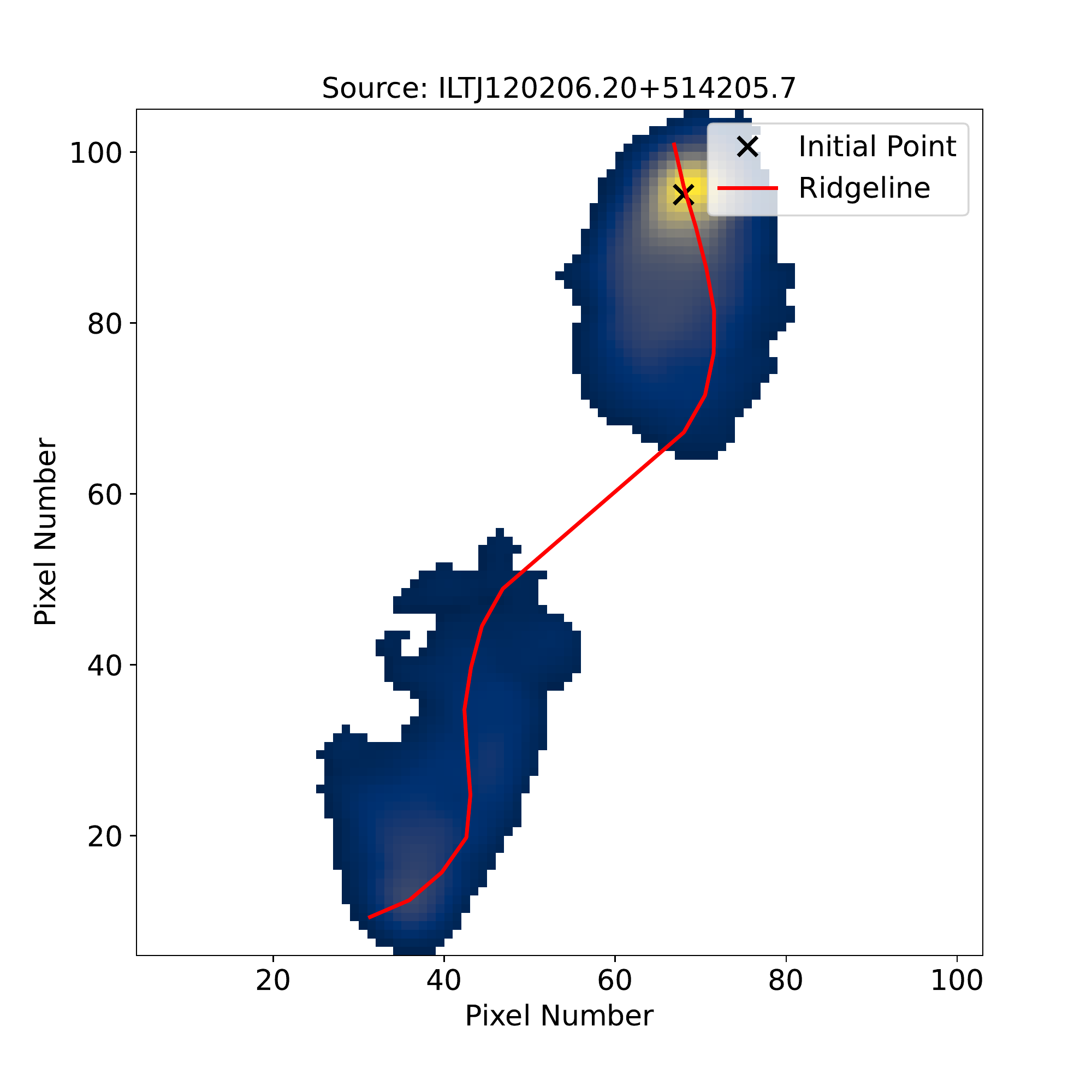}
        \caption{\label{fig:SucFRII}}
    \end{subfigure}
    \caption{The figure shows four examples of successfully drawn ridgelines (1 pixel $\approx$ 1.5 arcsec).  (a): A simple straight source with a ridgeline.  (b): A more complex, curved source demonstrating the ridgelines capacity to follow the path of the radio galaxy.  (c): A wide angled tail radio galaxy with the morphology clearly demonstrated with the ridgeline.  (d): An example where the ridgeline has successfully jumped from one part of the emission to another in an FRII.}
    \label{fig:Success}
\end{figure*}

\subsubsection{Initial Ridgeline Detection}
As the ridgeline is intended to trace the pathway of highest flux \rev{density} in the radio source, the \revb{maximum brightness} inside the source region is taken as the initial point. This maximum is calculated as the pixel inside the source region with the \revb{highest value}. This guarantees the ridgeline passes through the brightest point of the source, which in many cases will coincide with an AGN core or hot spot region, which we would also expect to lie along the ridgeline and so acts as a satisfactory starting point.  

The initial directions, in which RL-Xid searches for the ridgeline steps, are determined by finding the first two local maxima closest to the initial point.  If there is only a single maximum found then the complementary direction at $\pm\pi$ radians is taken (see Figure \ref{fig:Examples}).  If no direction can be determined, the initial point is masked, and a new maximum is located by taking the pixel with the next highest \revb{value} within the source regions. The erosion and direction-finding processes are iterated until starting directions are found or a maximum number of iterations is achieved (see Figure \ref{fig:flow}).  The initial directions form the centres of two initial search \revb{sectors}.  As given in Table \ref{tab:Params} the step size of the code, \texttt{R}, is set to 5 pixels.  This is roughly the same size as a beam’s width, with 1 pixel $\approx$ 1.5 arcsec and an angular resolution of 6 arcsec.  Further optimisation of \texttt{R} may improve the performance of the code in particular situations.  After exhaustive testing, a half \revb{sector} size was set as \texttt{$d\phi$} = 60 degrees. This produces optimal results between ridgelines that are too straight to be representative and those that, in some situations, tended to turn back towards the centre rather than continuing along the extent of the source.

\begin{table}
    \centering
    \caption{A list of the adjustable parameters in the code.  The name of the parameter is listed along with the current value used with LoTSS DR1, and a description of how it is used.  These parameters are likely to require adjusting according to survey specifications.}
    \begin{tabular}{*{2}{|c}|p{3.8cm}}
    \hline
    Parameter & Value (units) & Description \\
    \hline
    \hline
    \texttt{R} & 5 (Pixels) & Starting step size  between each ridge point. Set to LOFAR beam size in pixels. \\
    \texttt{$d\phi$} & 60 (degrees) & Half size of the search \revb{sector}. \\
    \texttt{rdel/ddel} & $\pm$ 0.000416667 & The degree to pixel {color{red} conversion factor}, currently set to the LoTSS configuration. \\
	\texttt{Rad} & 2.5 (Pixels) & Radius of the mask over an unsuccessful initial point determination.  Set to the resolution of the LoTSS beam.\\
	\texttt{ipit} & 6 & The number of iterations of reattempts on the initial point function. \\
	\texttt{Rmax} & 0.95 $\times$ source size & The total length permitted for each half of the ridgeline. \\
	\texttt{Jlim} & 0.4 $\times$ source size & The limit on the ridgeline length at which it is permitted for the algorithm to perform a 'jump'. \\
	\texttt{Jmax} & source size & The maximum distance to which a ridgeline can search when performing a 'jump'.\\
    \hline
    \end{tabular}
    \label{tab:Params}
\end{table}

\subsubsection{Ridgeline Tracing}
\label{SubSec:TR}
Having masked all but the initial search \revb{sectors}, an annular slice of the \revb{sector} is searched for a maximum \revb{brightness} value.  This slice is created by masking the previous point and any points greater than \texttt{R}, typically leaving only a segment of pixels, as shown in Figure \ref{fig:Examples}.  If the search area is empty, the search is repeated; increasing the width of the segment by increasing \texttt{R} one pixel at a time. This continues until either a value is found, or \texttt{R} has increased to be greater than \texttt{Rmax}; see Table \ref{tab:Params}.  \revb{The \texttt{Rmax} parameter, which sets the overall ridgeline maximum length, is set to a value related to the input catalogued source size, obtained as described in Section \ref{sec:Data}. The motivation for relating \texttt{Rmax} to source size is that we would expect for typical morphologies, allowing for source bends, the overall ridgeline length is unlikely to exceed the total source extent by more than a factor of $\sim$2.  As we generate the two ridgeline halves separately, and in some cases the starting point is not the source centre, there is a trade-off between the edge cases of (i) highly asymmetric sources that should not be restricted from completing their path and (ii) complex, more amorphous objects where too large a value of \texttt{Rmax} will lead to a ridgeline that continues to explore regions of lower dynamic range after it reaches close to the source edge, leading to 'looping' (see Appendix \ref{SubSec:Results}).}   Our strategy of increasing \texttt{R} where emission is not found within the search region allows the ridgeline to jump any nearby gaps in emission, for example in an FRII morphology where an emission-free region between the two lobes may be present; this is demonstrated in Figure \ref{fig:Success}.

Once a maximum point has been determined it becomes the starting point for the next step.  The previously identified point is recorded and masked out in a circle of radius \texttt{R}.  This prevents any future searches in the area from selecting the previous points or nearby points within a beam area.  A search \revb{sector} is created on the newly identified starting point, with a direction centred using \revb{the brightness-weighted average} of the previous annular search slice, see Figure \ref{fig:Examples}.  From here the point determination is repeated.  Again, if the search area is empty, \texttt{R} is increased until a new point can be found (i.e. the ridgeline is permitted to `jump' empty regions), or restrictions are met.  Here the length is restricted again to \texttt{Rmax}; however, the jump is only permitted if the ridgeline is under \texttt{Jlim} and must not exceed \texttt{Jmax} in total, see Table \ref{tab:Params} for definitions and values.  This helps prevent the ridgeline from extending to any possible remaining external un-associated emission.

Similarly to \texttt{Rmax}, the values of \texttt{Jmax} and \texttt{Jlim} were determined through a series of tests and visual checks for quality, and will need to be adjusted for different surveys.  These values optimized the successful outcomes and the quality of the ridgelines by reducing issues such as extending to nearby sources, and maximizing the number of FRII jumps achieved.

After the first point in each direction has been determined, the process iterates until ridge points can no longer be located.  The ridgeline is completed when the restrictions are met, or when no viable points are found within the search \revb{sector} , as shown in Figure \ref{fig:flow}.  The final step outputs the information about the ridgeline's individual points as a text file.  This includes the location of the ridge points in pixels, the angular direction of the ridge points in radians, and the length of the ridgeline. An image of the source with an overlay of the ridge points plotted and connected (see Figure \ref{fig:Success}), is saved for visual comparison.  Using Figure \ref{fig:Examples}  the complete ridgeline definition process can be summarised as:

\begin{enumerate}
    \item Initial point is found (black cross)
    \item Initial direction and search \revb{sectors} are located - \textit{If not, source flagged as a \textit{\revb{Failed}}}
    \item Initial points are determined - \textit{If not, flagged as a \textit{\revb{Failed}}}
    \item Search \revb{sectors} are placed at the first ridge points in the direction from the initial point (magenta triangles are examples of how the \revb{sectors} appear)
    \item 	New ridge points are determined (cyan dots) - \textit{If the search area is empty the step size can increase to allow for bridging of gaps in emission}
    \item Previous points are masked around, with a radius one step size (white circles)
    \item The most recent ridge points are used for search \revb{sectors}
    \item Repeat steps (iv) to (vii) until the end of the source is reached, or the ridgeline is greater than \texttt{Rmax}
    \item Ridgelines are completed for both directions
    \item The information about the ridgeline is printed to a text file and an image of the source is produced with the ridge points (cyan dots) joined together (red arrows), leading out from the initial point (black cross) overlaid on it.
\end{enumerate}

Throughout the ridgeline creation process there are steps where the code can fail and output a \textit{\revb{Failed}} source, see Figure \ref{fig:flow}.  During the \texttt{FloodFill} process, cataloguing errors and mismatching component information causes the sources to fail.  There is a chance a source may fail the \texttt{Erosion} process, mainly due to rare catalogue errors creating empty source images.  The \textit{\revb{Failed}} source is designated as having an \textit{ID Out of Region} when these catalogue errors produce incorrect cut-outs; the initial point is deemed to be outside the initial search area.  Assuming no cataloguing errors occur, if the initial point iterations exceeds the \texttt{ipit} parameter (see Table \ref{tab:Params} for definition and value) without finding initial directions then the source has failed due to lack of suitable initial points.  Likewise, when searching for the first points, if a value is not found within \texttt{Rmax} the source is classed as a \textit{\revb{Failed}}.  This may occur where the initial point is on one lobe and the diagonal distance to the other lobe is greater than \texttt{Rmax}, or where the initial point is on the edge of the source and there is no first point to be found in one direction.  Both first points must be found for a source to continue to Completion.

\subsection{Results}

This ridgeline drawing process has two possible outcomes: \textit{\revb{Completed}}, where a ridgeline has been drawn, or \textit{\revb{Failed}}.  These outcomes are discussed in detail in Appendix \ref{SubSec:Results}.  For 95.9 per cent of our sample the code \textit{\revb{Completed}}.  This generates output files containing the numerical and graphical information.  Examples of \textit{\revb{Completed} sources can be seen in Figure \ref{fig:Success}}, where a variety of ridgelines are shown demonstrating how they can pick out simple, straight structures in Figure \ref{fig:SucRL}, to more complex, angled sources in Figure \ref{fig:SucBent}.  This complexity can lead to distinguishing different morphologies such as wide angled tail (Figure \ref{fig:SucWAT}) and narrow angled tail sources.  The\textit{ \revb{Completed}} dataset are described in detail in Appendix \ref{subsubsec:comp}; and the \revb{\textit{Failed}} outcomes discussed in more detail in Appendix \ref{subsubsec:fails}.

\newlength\imagewidth
\newlength\imagescale

\begin{figure*}
    \centering
    \begin{minipage}{0.5\textwidth}
    \pgfmathsetlength{\imagewidth}{\linewidth}%
    \pgfmathsetlength{\imagescale}{\imagewidth/524}%
    \begin{tikzpicture}[x=\imagescale,y=-\imagescale]
        \node[anchor=north west] at (0,0) {\includegraphics[width=\imagewidth]{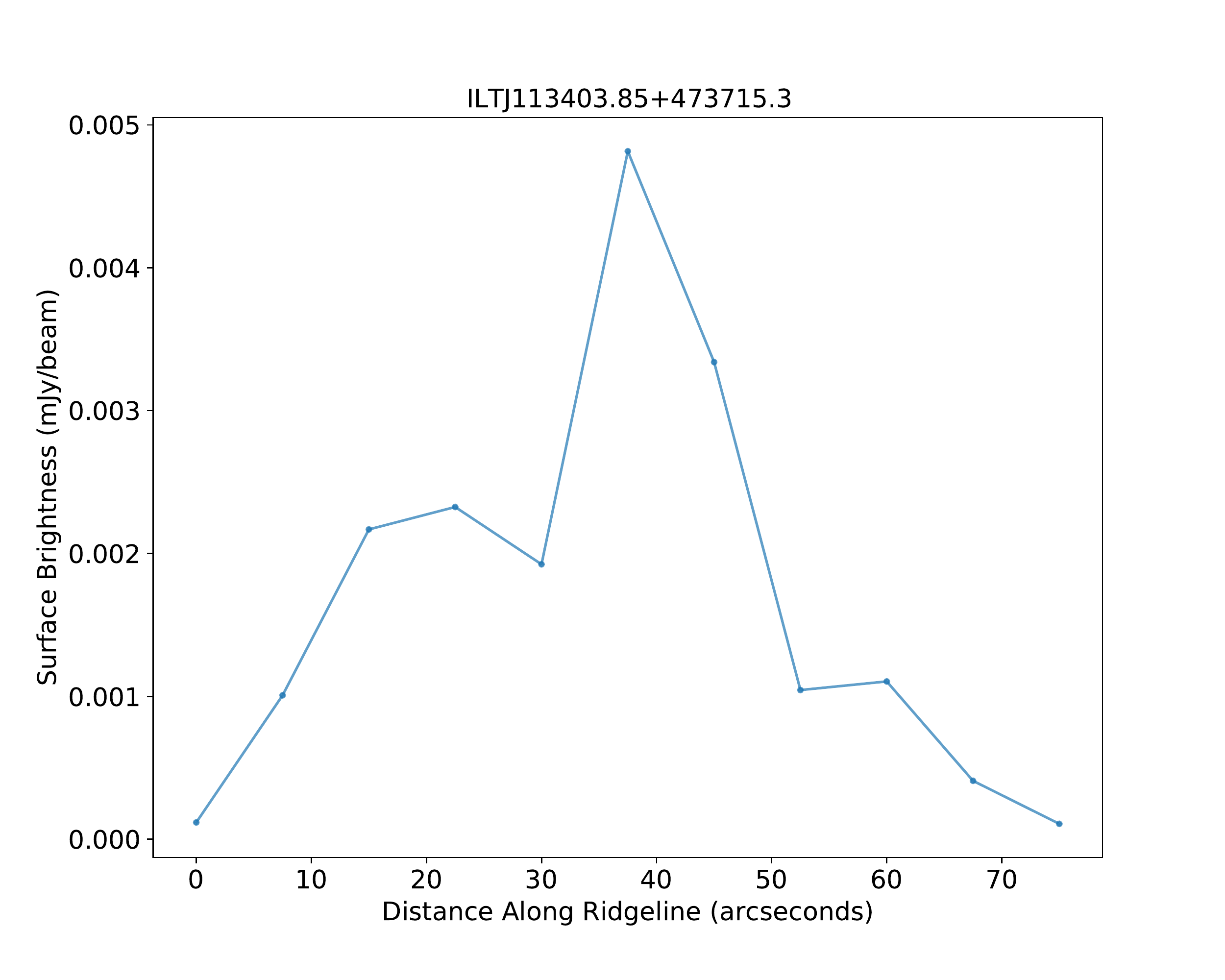}};
        \node[anchor=north west] at (325,60) {\includegraphics[width=0.275\imagewidth]{Ridgelines/WAT.pdf}};
    \end{tikzpicture}
    \subcaption{\label{fig:SBPPEak}}
    \end{minipage}%
    \begin{minipage}{0.5\textwidth}
    \pgfmathsetlength{\imagewidth}{\linewidth}%
    \pgfmathsetlength{\imagescale}{\imagewidth/524}%
    \begin{tikzpicture}[x=\imagescale,y=-\imagescale]
        \node[anchor=north west] at (0,0) {\includegraphics[width=\imagewidth]{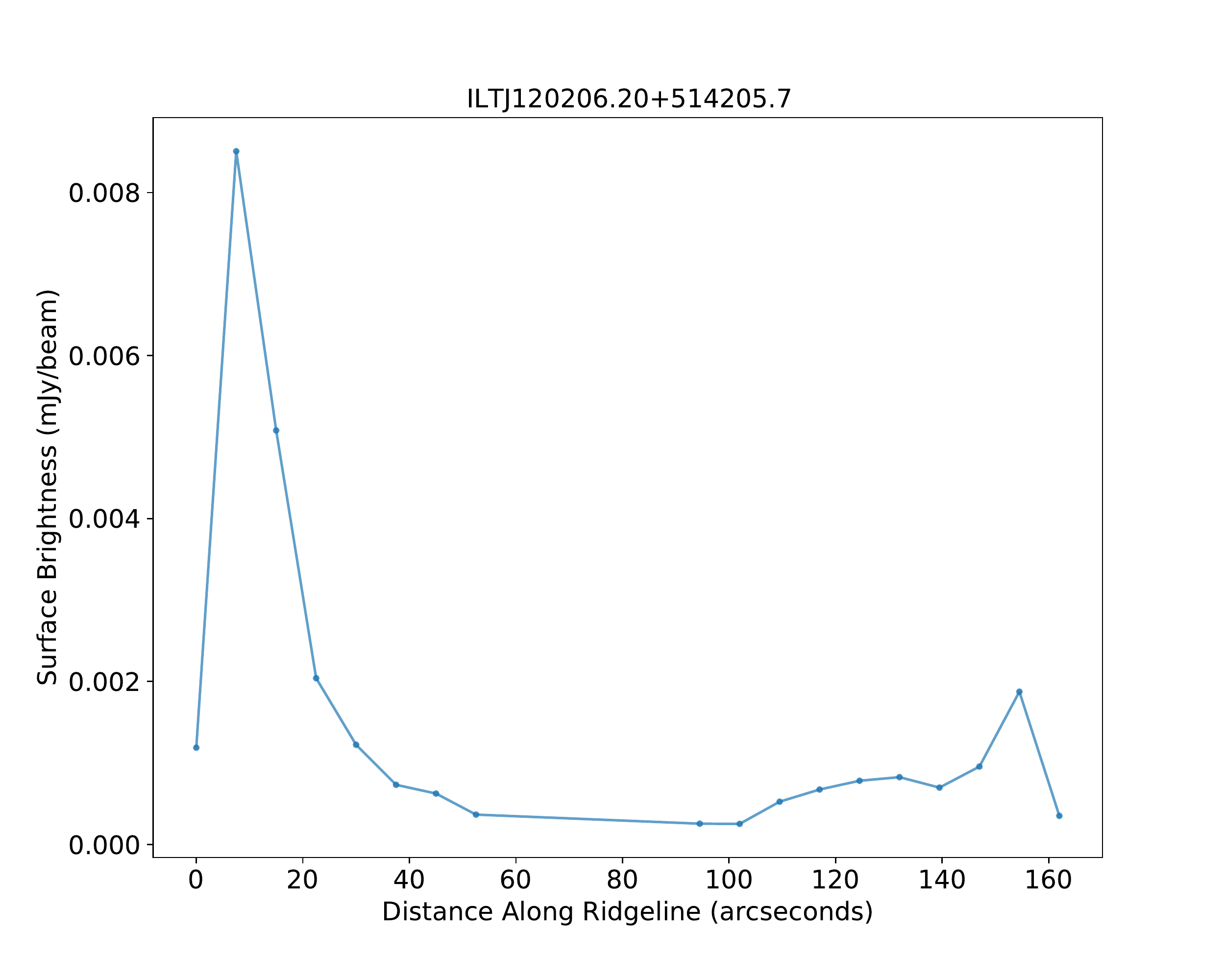}};
        \node[anchor=north west] at (325,60) {\includegraphics[width=0.275\imagewidth]{Ridgelines/FRII.pdf}};
    \end{tikzpicture}
    \subcaption{\label{fig:SBPDip}}
    \end{minipage}
    \caption{Two surface brightness profiles with the corresponding ridgeline in the top right insert.  \rev{(a): The surface brightness profile for a wide angle tail (FRI). (b): The surface brightness profile for an FRII.}  The profiles show the surface brightness for the distance along the ridgeline from one end. For the left hand image the starting position on the SB profile corresponds to the upper leftmost end of the ridgeline, and in the right hand image the starting position corresponds to the upper rightmost end of the ridgeline.}
    \label{fig:SBProfiles}
\end{figure*}

\section{Surface Brightness Profiles}
\label{sec:SB}

In addition to recording the positional information, RL-Xid also records a surface brightness profile along the ridgeline, which may be used as part of the optical identification process and for morphological classification.  The surface brightness (SB) profile is the measurement of SB of the radio source at each point along the ridgeline.  A \revb{brightness-}weighted average of nearby pixels is plotted for a graphical representation of the profile of the ridgeline.  The SB profile shows the dips and peaks in \revb{brightness} along the ridgeline.  Using the definition of FR classification (described in the introduction) it is expected that sources with FRI type morphology will have SB peaks near the centre of their ridgelines.  Likewise, those with an FRII type structure will have two distinct peaks towards the ends of the ridgeline, demonstrating the difference between FRI and FRII sources, see Figure \ref{fig:SBProfiles}.  

\subsection{Categorisation of the Surface Brightness Profiles}

In order to look at the SB profiles in more detail and use them for host identification and morphological classification, the profiles were classified according to the location of any dips or peaks.  The SB profiles were split into four different groups depending upon the location of a minimum value along the ridgeline: \textit{Dips}, \textit{Peaks}, \textit{Both} and \textit{Neither}.  We considered the optimally sized central region within which a peak or dip is expected to fall, and found the central 30 per cent of the source gave the best results.  \rev{This method differs from the original Fanaroff-Riley classification \citep{Fanaroff1974} where they excluded any compact component situated on the central galaxy, and calculated their ratio from this central location.}

If only a minimum value occurs within the middle 30 per cent of the length of the ridgeline then the SB profile is classed as a \textit{Dip}.  In order to determine if a minimum value is contained in the middle 30 per cent the lowest three values of the SB profile are taken.  As the dips are meant to be global minima which are also not the minima associated with the ends of the ridgeline, we start with the lowest; the first step is to check to see if it lies at the end of the ridgeline.  This is expected, as the ridgeline will commence or terminate at the edge of the source, possibly in an empty part of the array.  If the lowest point is at the end of the ridgeline the next lowest point is chosen and checked.  The three points are checked in ascending order until one point not at the end of the ridgeline is found.  The location of this point along the ridgeline is compared with the middle 30 per cent of the ridgeline.  If it lies in this region the source is classed as having a dip.

The ridge point with the highest associated SB is labelled the peak and its location is checked against the middle 30 per cent of the ridgeline.  If only the maximum value and not the minimum value is in this region the profile is designated a \textit{Peak}.  For those profiles where both a dip and a peak are found in the central region they are classed as \textit{Both}, and if neither are found they are classed as \textit{Neither}.

\begin{table}
    \centering
    \caption{The table shows the percentage of surface brightness profiles which are classified as either an FRI, FRII or Hybrid, in each of the \textit{Dip} and \textit{Peak} groups.}
    \begin{tabular}{*{3}{|c}|}
    \hline
    ~ & Dips (\%) (Number) & Peaks (\%) (Number) \\
    \hline
    \hline
    FRI & 16 (35) & 71 (262) \\
    FRII & 59 (129) & 11 (40) \\
    Hybrid & 25 (56) & 18 (67) \\
    \hline
    \end{tabular}
    \label{tab:DipPeak}
\end{table}

\subsection{Comparison with LoMorph}
\label{subsec:ComLoM}

As the data are a subset of the sample from \cite{Mingo2019} this method of sub-dividing the SB profiles could be compared to LoMorph's classification of these radio sources to investigate the potential of ridgelines for morphological identification.  \revb{It should be noted the LoMorph's classification excludes any compact component situated on the central galaxy.}  The LoMorph classification of the sample sources as \textit{Star-Forming}, \textit{Double-double}, \textit{Fuzzy blobs}, \textit{Core-bright} and \textit{Bad} were left out, leaving a clean sample (88 per cent of the SB profiles) of FRI (including wide angle tail (WAT) and narrow angle tail (NAT)), FRII, and Hybrids.  This was done as a starting point to assess the capabilities of RL-Xid to perform simple morphological classifications.  The final sample from \cite{Mingo2019} consists of 2106 sources of which $\sim$60 per cent are FRI, $\sim$20 per cent FRII and $\sim$20 per cent are Hybrid.  In comparison the final SB groups show similar percentages across the \textit{Dip} and \textit{Peak} groups (see Table \ref{tab:DipPeak}).

\begin{figure}
    \centering
    \includegraphics[width=\linewidth]{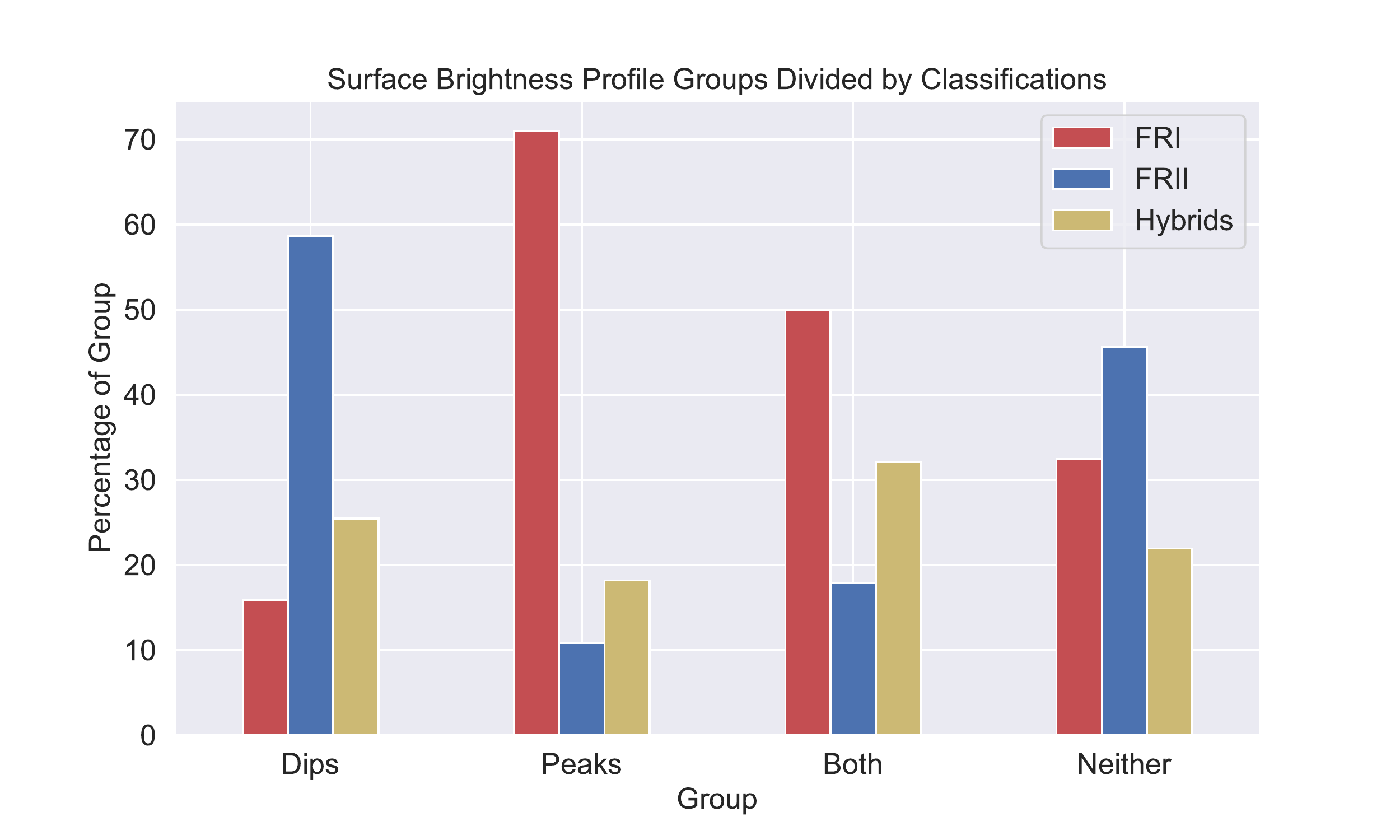}
    \caption{The four surface brightness groups separated into by-eye classifications. This shows the percentage of each group, defined by the presence of a peak or dip in the central 30 per cent of the ridgeline, matching each LoMorph classification.}
    \label{fig:Class}
\end{figure}

Figure \ref{fig:Class} shows a high number of FRI sources having SB profiles in the \textit{Peak} group and the majority of the \textit{Dip} SB profiles as matching sources with FRII classification.  As WATs and NATs are subsets of FRIs \citep{Owen1976, Rudnick1976, Odea1985, Hardcastle2020} it is expected that they make up a high number of the \textit{Peak} SB profiles.  Likewise, as FRIIs and hybrids are thought to have similar morphologies \citep{Mingo2019, Harwood2020} the \textit{Dip} SB profiles will contain a high number of these classifications.  Since Hybrids are a heterogeneous population \citep{Mingo2019}, unsurprisingly they do not fit neatly into any of the categories.  This is the most likely cause of the high number of Hybrids in the \textit{Both} SB profile group where both a peak and a dip were found in the central region.  Likewise, some FRIIs have a structure with a bright central core and emission gaps between lobes; this will increase the number of FRII classified sources likely to appear in the \textit{Both} SB profile group.  Interestingly, the \textit{Neither} SB profile group has a higher FRII content.  This group appears to be made up of sources with a fairly uniform surface brightness, so unambiguous classification by any method is challenging.  Table \ref{tab:DipPeak} and Figure \ref{fig:Class} demonstrate a link between the presence of a dip or peak in the middle 30 per cent of a ridgeline and the morphology of a source.

In addition to using the SB information for morphological classification, we also wanted to explore its use in host identification.  For this purpose, we identified a location that characterised the most likely host location given the surface brightness profile.  Taking the \textit{Dip} SB profiles, which can be considered to have an FRII like morphology with two outer peaks, the ridge point corresponding to the location of the dip was recorded.  All other sources recorded the ridge point at which the maximum occurred.  For the \textit{Peak} SB profiles, this was because of their FRI like morphology coinciding with a central peak; for the \textit{Both} SB profiles this included those with FRI like tendencies and those FRIIs with bright central cores.  For the \textit{Neither} SB profiles, the peak was taken as a starting point to work with.  \rev{In each group the} distance between the host galaxy for this source and the corresponding dip or peak is considered to be the SB separation used in later calculations.

In conclusion, we have shown that RL-Xid is extracting useful morphological information.  Future work is planned to develop this capability of the code further to enable more sophisticated classifications.

\section{Incorporating Ridgelines into Host Identification}
\label{sec:LR}

\revb{Cross-identification for extended sources is an inexact science and automated methods are becoming increasingly necessary to handle large volumes of data containing increased numbers of extended or multi-component sources.  The aim of our work is to develop the optimal likelihood ratio method for complex extended sources, incorporating spatial information from the ridgelines as well as existing catalogue information.  This section compares how well several different definitions of the host-galaxy/radio source positional offset parameter (hereafter called "separation", and defined for each case considered in Section \ref{SubSec:SP}) perform within a maximum likelihood approach applied to extended sources whose hosts were originally originally matched through LGZ.  Building directly on \cite{Williams2019}'s work, “radio” refers to the PyBDSF radio source catalogue (LoTSS) and “optical” refers to the matching catalogue (PanSTARRS/AllWISE), even though it contains IR sources.}

\citet{Williams2019} used two methods to determine the optical/IR counterparts of a radio source.  The first is a likelihood ratio (LR) which identifies the counterpart in a statistical fashion.  Based on work by \cite{Richter1975, DeRuiter1977} and \cite{Sutherland1992} the ratio of the probability of a source being a true counterpart to it being a random interloper is used to statistically investigate whether an object at one wavelength is the correct counterpart to an observed object at another wavelength.  \cite{Williams2019} considered magnitude, colour and distance from the radio source and its potential counterpart, taking into account the uncertainties of each, and selected the host based on the LR for all radio sources < 30 arcsec in size and all PyBDSF components < 30 arcsec in size which may have been incorrectly combined into larger sources.    

Out of the $\sim$320,000 sources in the LoTSS catalogue $\sim$12,000 went to LGZ for visual classification \citep{Williams2019}.  Of the 950 sources analysed in this paper and classified as \textit{\revb{Completed}}, 99.8 per cent (948) had their optical ID's determined via LGZ.  In the following sections these 950 sources have been used to investigate our adaption of the LR method described by \cite{Williams2019} to incorporate ridgeline information.

\subsection{Separation Parameters}
\label{SubSec:SP}

\cite{Williams2019} use the separation between \rev{the flux density} weighted LOFAR catalogue radio position and the position of the possible optical counterpart.  This method assumes the true radio and optical source positions are the same, and any separation between the radio source and its counterpart is due to statistical uncertainties.  In what follows we use the "Centroid distance" to refer to this separation measure; however, as discussed in Section \ref{subsec:SPF}, we relax the underlying assumption regarding the statistical origin of deviation between host and radio centroid so as better to reflect the nature of extended source behaviour.  We use Centroid distance as a comparison to our ridgeline separation measure in the analysis which follows.

The "Ridge distance" is the shortest separation between a possible counterpart and the ridgeline.  It is the perpendicular distance determined using the line segment joining the nearest two consecutive ridge points to the possible counterpart.  If the perpendicular line from the counterpart lies outside the line segment on the ridgeline, then the distance to the closest ridge point is used.  Similarly to the original method the assumption is that within uncertainties the true host should lie on the ridgeline.  The SB separation, consisting of the distance of the host to the morphological central feature, i.e. peak or dip, as described previously (see Section \ref{subsec:ComLoM}) is the third distance parameter which can be considered.

We carried out preliminary testing using only the radio parameters to investigate the performance of the three separation parameters.  This demonstrated the promise of using a ridgeline-based distance measure and also showed that combining with a radio centroid-based measure showed potential.  Our tests showed that whilst the ridge distance performed comparably to the the centroid distance, the best results were obtained by taking the geometric mean of the two, rather than just taking the possible counterpart with the highest LR using either separation.  The SB distance measure performed less well and so has not been applied further in this work; however, its use could be further explored after future code refinements such as smoother profiles and improved peak detection.

\subsection{Application of Magnitude and Colour}
\label{sec:MC}

The combined optical catalogue of Pan-STARRS and AllWISE sources has had no pre-filtering for astronomical objects such as star forming galaxies, or stars.  Radio jets are known to be associated with hosts of particular colours and magnitudes, leading \cite{Williams2019} to apply these parameters in their LR.

The LR \cite{Williams2019} used to statistically investigate whether a source observed at one wavelength is the correct counterpart (or host) to a source in another wavelength, is calculated as the ratio of the probability of the source being the correct counterpart over being a random source. This is given by:
\begin{equation}
    \centering
    LR = \frac{q(m, c)f(r)}{n(m, c)}
    \label{eq:LR}
\end{equation}
where $q(m, c)$ is the a priori probability of a radio source having a counterpart of magnitude, $m$, (split into fixed colour bins, \rev{c}) and $n(m, c)$ is the sky density per unit area (arcsec$^2$) of objects at this magnitude (split into the same colour bins, \rev{c}). $f(r)$ is the probability distribution of the offset \citep{Sutherland1992, Williams2019}.  \revb{As with \cite{Williams2019}, the following methods were carried out using PanSTARRS i-band and AllWISE W1 magnitudes and the resulting i - W1 colour.}

\subsection{Calculating $q(m,c)$ and $n(m,c)$}
In order to determine the probability distribution which reflects the true distribution of host properties three alternative \rev{routes} were considered for $q(m,c)$ and tested using \rev{only the} magnitudes. 

\begin{enumerate}
    \item The first method investigated was the original technique by \cite{Williams2019}.  This searches for sources with matching magnitudes across all nearby sky patches, for all given radio sources.  However, \cite{Williams2019} were aiming to cross-match to the full radio population and the extended sources being studied here may well have a different range of properties.
    \item The second distribution \rev{was created by selecting the closest optical source to each} LOFAR catalogue position.  These were selected because $\sim$50 per cent of these are known to be true counterparts, from checking with LGZ.  This can help counter any biases from using only the known hosts as it is an almost even divide between both counterparts and the whole population. 
    \item The final distribution was generated from the population of 950 known hosts, as this is the best representation of the host properties of complex, extended sources classified by LGZ.  This is likely to give overly good results when applied to the same population, but better reflects the most appropriate approach to take with future LOFAR survey sky areas where we can make use of the improved knowledge developed with LoTSS DR1.
\end{enumerate}

For \rev{(ii) and (iii)} kernel density estimators (KDEs) \rev{\citep{Pedregosa2011}} were generated (Gaussian kernel, bandwidth = 0.2, bin size = 0.05) for both the i-band and the W1-band.  Out of all three methods, using the $q(m)$ \rev{in (iii) very slightly outperforms the other two methods by at most less than 0.5 per cent}, and so we adopted option (iii) as this best represents the population of extended sources.  \rev{Applying colour, a} 2D KDE (Gaussian kernel, bandwidth = 0.2, bin size = 0.05) was generated for the known hosts of DR1 in colour and magnitude for both the i-band and W1 band and were used as $q(m,c)$.

Due to the large number of sources in the optical catalogue a random sample of 50 000 were selected to form a 2D KDE (Gaussian kernel, bandwidth = 0.2, bin size = 0.05) for colour and magnitude in both bands for $n(m,c)$.  Multiple samples were run and checked to make sure they maintained the same properties.

\begin{table*}
    \centering
    \caption{The percentage of correctly identified hosts in the sample sets obtained using the Centroid and Ridge separations, the combination of both, and the union of both.  The results are split into three sections: the original large and bright group (\textit{Left most columns}) those with a size over 60 arcsec and flux density greater than 30 mJy; the full DR1 sample (\textit{Right most columns}) everything over 15 arcsec and 10 mJy; and those of an intermediate size and flux density (\textit{Centre two columns}).  The results are given first a a percentage of successes from the number of successfully drawn ridgeline sources in each category, and secondly as a percentage of all the sources in the category.}
    \begin{tabular}{*{7}{|c}|}
    \hline
     ~ & \multicolumn{2}{c}{S > 30 mJy and $\theta$ > 60 arcsec (950 Sample)} & \multicolumn{2}{c}{10 < S < 30 mJy and 15 < $\theta$ < 60 arcsec} & \multicolumn{2}{c}{S > 10 mJy and $\theta$ > 15 arcsec (Full DR1 AGN)} \\
    \hline
     Separation & \% Ridgelines & \% Group Total & \% Ridgelines & \% Group Total & \% Ridgelines & \% Group Total  \\
    \hline
    \hline
    Centroid & 89.8 & 86.1 & 97.8 & 80.6 & 95.6 & 82.0  \\
    Ridge & 91.5 & 87.7 & 96.8 & 79.7 & 95.3 & 81.7 \\
    Combined & 92.4 & 88.6 & 98.0 & 80.7 & 96.4 & 82.7  \\
    Union & 95.5 & 91.5 & 99.0 & 81.6 & 98.0 & 84.1  \\
    \hline
    \end{tabular}
    \label{tab:Results}
\end{table*}

\subsection{The Separation Probability Function}
\label{subsec:SPF}
\rev{\cite{Williams2019} define $f(r)$ as} the probability distribution of the offset between the optical source and the flux \rev{density} weighted LOFAR catalogued position.  We have taken $f(r)$ to have the form:
\begin{equation}
    \centering
    f(r)\ =\ \frac{1}{{2\pi\sigma}^2}{e}^\frac{{-r}^2}{{2\sigma}^2}.
    \label{eq:FR}
\end{equation}

Using the previous distance-based investigations the best results were achieved through taking the geometric mean of the $f(r)$ of the centroid distance and the ridgeline distance.  We therefore tested the magnitude and colour LR formulation with the outcomes of $f(r)$ for each of the ridgeline and centroid distance definitions separately.  We note the formally correct distribution for the ridgeline distance parameter is a 1D rather than 2D Gaussian.  The method performance for the Ridge distance is not found to differ significantly if a 1D or 2D Gaussian distribution is used.

For the Ridge separation, maintaining the assumption that the host will lie on the ridgeline within positional uncertainties gives $\sigma^2 = \sigma_{\rm{rad}}^2 + \sigma_{\rm{opt}}^2 + \sigma_{\rm{ast}}^2$.  The astrometric uncertainty between the optical and radio catalogues, was chosen to be $\ \sigma_{\rm{ast}} = 0.6$ arcsec, from \cite{Williams2019}; the optical postional uncertainty for each source, $\sigma_{\rm{opt}}$, is taken from the optical catalogue and, after extensive testing, we took $\sigma_{\rm{rad}}$ to be 3 arcsec (2 pixels).  The choice of $\sigma_{\rm{rad}}$ is related to the step and \revb{sector} size (\texttt{R} and \texttt{$d\phi$}), and is the optimal setting for the LR given the systematic uncertainties producing some of the issues discussed in Appendix \ref{SubSec:Results}.

For the Centroid separation, as $\sigma$ represents the assumed width of the distribution of the separation, $r$, we used the known distribution of catalogue/host offsets for extended sources from LOFAR DR1 to derive empirically that $\sigma = 0.2$ in units of the source size (i.e. the host is typically within the central 40 per cent of the radio source extent).  This empirically derived distribution better accounts for the known broad distribution of the centroid position host offsets, with the centroid separation taken as the distance from the possible counterpart to the LOFAR catalogue position as a fraction of the size of the source as given by \cite{Mingo2019}.  Our method emphasises the possible counterparts which are within the size of the source, whilst reducing the impact from those further out.

As we are matching this set with the known hosts from LGZ, all of which have AllWISE identifications, we chose to use only the W1 magnitude as a parameter in Equation \ref{eq:LR}.  LGZ found many instances where an AllWISE source was present in the expected location of a host, particularly for small double radio sources, but no PanSTARRS source was present.  We have chosen to use the W1 magnitude because it is more likely that a counterpart will exist in this band.  This may need adjusting for future datasets where there is the possibility of detection in different bands.

\subsection{Outcome of the Likelihood Ratio}
\label{subsec:OutLR}
The results from this analysis are shown in \ref{tab:Results}, and it is clear both separation parameters perform well individually.  The maximum number of correct hosts which can be found, for the large and bright sources (S > 15 mJy $\theta$ > 60 arcsec), is 95.5 per cent of the ridgeline sample or 91.5 per cent of the full group.  This number is obtained via the union of both separations, with 816 correctly identified by both separations, 38 correctly identified by the centroid separation alone and the 54 identified by the ridgeline separation alone.  This demonstrates that automated methods of source identification can be very successful for extended sources.  Table \ref{tab:Results} demonstrates that the combination of the ridgeline and centroid, where the geometric mean of the distance probability functions was taken before the LR was calculated, successfully identified 92.4 per cent of hosts.  For future datasets in which the true hosts are not known, this method is most suitable as it does not need prior information about which distance parameter will perform best.

Having established the success of the cross-identification method, we tested it on a wider range of flux \rev{density} and source sizes.  RL-Xid was applied to all sources in the \cite{Hardcastle2019} AGN catalogue which satisfy the criteria of a flux density > 10 mJy and a size > 15 arcsec.  There are 3964 sources meeting the size and flux \rev{density} requirements, of which \rev{RL-Xid successfully drew ridgelines for 3384 sources}.  The LR was then calculated for these 3384 sources using the different separations and colour and magnitude information in the same way as for the large, bright source sample.  The results are given in Table \ref{tab:Results}.  We use the union of the results from the different separation parameters in conjunction with the host galaxies properties to produce a success rate of 98.0 per cent of correct host galaxies found.  This can be attributed to the slightly smaller sized sources than the original sample having a smaller population of nearby galaxies from which to select possible counterparts. This is demonstrated in the centre two columns of Table \ref{tab:Results}, where the number of hosts found, as a percentage of the ridgelines drawn, is 99.0 per cent in the AGN catalogue.  These are the intermediate sources which satisfy the criteria of having a flux \rev{density} between 10 and 30 mJy and a size between 15 and 60 arcsec. 

As this method will be applied to datasets with no prior hosts determined, we also compare the combined Centroid and Ridge separation results directly with the performance of the previous host identification methods from \cite{Williams2019} for DR1. In the full DR1 sample of 3946 sources, 33.7 per cent (1331) were previously identified using \cite{Williams2019}'s original LR.  Of the 3384 where the ridgelines successfully drew, 28.7 per cent (971) were already identified through their LR method. For the 3263 where the hosts were correctly identified through RL-Xid, 29.6 per cent (965) were already known. For the remaining 121 unidentified ones, 5.0 per cent (6) were already identified.  In summary RL-Xid was able to identify an additional 58.2 per cent (2298) hosts in the selected full AGN sample compared to the LR in \cite{Williams2019}.

Of the 562 AGN meeting the size and flux \rev{density} requirements, for which ridgelines could not be drawn, most failed due to catalogue errors or the sources being circular and centre bright.  This structure is ideal for identification through the original LR method and indeed 64.1 per cent (360) of these hosts have already been identified in this way, and it is therefore unlikely that these sources would need to be identified using RL-Xid.  For the remaining sources where a ridgeline could not be drawn, using the centroid is a viable method for determining counterparts.

\section{Summary}
\label{sec:Sum}

In LoTSS DR1 a likelihood ratio method was used to determine the majority of IR/optical counterparts.  This statistical determination works very well for small, compact sources.  However, for larger, extended sources or those with a more complicated structure, for example, where blending of multiple sources may occur, the citizen science project LOFAR Galaxy Zoo was used to determine the host galaxies.

As an alternative to this labour-intensive process we have developed RL-Xid to draw ridgelines and perform cross-identification for extended radio sources. This method requires a given radio catalogue with correctly associated radio components for the non-single component sources.  In the full DR1 sample $\sim$76 per cent (2984) of the sources were considered to be single component sources, therefore at least 25 per cent of sources would have to be associated.  In this paper we demonstrate:

\begin{enumerate}
    \item RL-Xid was able to draw ridgelines on 85.8 per cent of the LoTSS DR1 sample of sources over 10 mJy and 15 arcsec, including 95.9 per cent of the sources over 30 mJy and 60 arcsec.
    \item Of the ridgeline sample group for the largest and brightest sources (over 30 mJy and 60 arcsec), RL-Xid was able to correctly identify 95.5 per cent of the hosts, using the union of the Centroid and Ridge distance parameter results for the likelihood ratio function.  For the full LoTSS DR1 ridgeline sample, using the same method, 98.0 per cent were correctly identified.
    \item The fainter (between 10 and 30 mJy) and smaller (between 15 and 60 arcsec) source subset had 99.0 per cent hosts correctly identified, mostly likely due to a smaller possible optical population surrounding the source.
    \item We have shown that the most successful method to be applied to a new dataset for which hosts are unknown is the combined Centroid and Ridge probability distribution function for separation: the likelihood ratio found 82.7 per cent of the full DR1 sample hosts, where the previous likelihood ratio method had found 33.7 per cent.  RL-Xid identified 49.0 per cent more hosts, significantly reducing the number of possible sources requiring cross-identification via LGZ, though association will still need to be performed.  This proportion could be increased by applying the centroid distance parameter only in cases where a ridgeline could not be drawn.
    \item Preliminary results demonstrate the effectiveness of surface brightness profiles as a complementary method of automated morphological classification which is well matched to the outcomes of LoMorph \citep{Mingo2019}.
\end{enumerate}

This method shows the potential of automated methods for cross-identification to be applied to extended sources.  The intention is for this method to be used in conjunction with the already existing LR method for DR2 of LoTSS to help identify IR/optical counterparts and reduce the number of sources requiring classification via the public LGZ.  The testing of RL-Xid was carried out on data with known hosts and pre-associated sources, so minor improvements were applied when run on the LoTSS DR2 dataset.  This includes a LR threshold, chosen to reject objects with no visible host, and code performance improvements, as more information about the true distribution of the host properties is incorporated.  Preliminary testing of RL-Xid on the LOFAR DR2 is already showing good performance results and as mentioned in Section \ref{sec:Data} pre-filtering for AGN does not appear necessary.  

This method can be applied to data from surveys other than LOFAR, for example it is currently being applied to a sample of MeerKAT data.  As well as being used for identifying host galaxies, RL-Xid has shown the ability to be a useful tool in helping to automate other processes such as morphological classification through SB profiles.

\section*{Acknowledgements}
We thank the anonymous referee for helpful comments that have improved the paper.  We thank Ken Duncan for his useful comments.
BB acknowledges a studentship from the UK Science and Technology Facilities Council (STFC). JC acknowledges support from the UK Science and Technology Facilities Council (STFC) under grants ST/R00109X/1 and ST/R000794/1.  BM acknowledges support from the UK STFC under grants ST/R00109X/1, ST/R000794/1, and ST/T000295/1.  PNB is grateful for support from the UK STFC via grants ST/R000972/1 and ST/V000594/1.  MJH acknowledges support from the UK STFC (ST/R000905/1).  JS is grateful for support from the UK STFC via grant ST/R000972/1.  WLW acknowledges support from the CAS-NWO program for radio astronomy with project number 629.001.024, which is financed by the Netherlands Organisation for Scientific Research (NWO).

LOFAR, the LOw Frequency ARray designed and constructed by ASTRON, has facilities in several countries, which are owned by various parties (each with their own funding sources), and are collectively operated by the International LOFAR Telescope (ILT) foundation under a joint scientific policy. The ILT resources have benefited from the following recent major funding sources: CNRS- INSU, Observatoire de Paris and Universit\'e d’Orl\'eans, France; BMBF, MIWF-NRW, MPG, Germany; Science Foundation Ireland (SFI), Department of Business, Enterprise and Innovation (DBEI), Ireland; NWO, The Netherlands; the Science and Technology Facilities Council, UK; Ministry of Science and Higher Education, Poland. Part of this work was carried out on the Dutch national e-infrastructure with the support of the SURF Cooperative through grant e-infra 160022 and 160152. The LOFAR software and dedicated reduction packages on \url{https://github.com/apmechev/GR} ID LRT were deployed on the e-infrastructure by the LOFAR einfragroup, consisting of J. B. R. Oonk (ASTRON \& Leiden Observatory), A. P. Mechev (Leiden Observatory) and T. Shimwell (ASTRON) with support from N. Danezi (SURFsara) and C. Schrijvers (SURFsara).

This research made use of ASTROPY, a community-developed core PYTHON package for astronomy \citep{Collaboration2013, Price-Whelan2018} hosted at \url{http://www.astropy.org/}, of MATPLOTLIB \citep{Hunter2007}, of NumPy \citep{VanDerWalt2011}, of \texttt{scikit-learn.KernelDensity.Neighbors} \citep{Pedregosa2011} hosted at \url{https://scikit-learn.org/stable/modules/generated/sklearn.neighbors.KernelDensity.html\#sklearn.neighbors.KernelDensity} and of Pandas \citep{Mckinney2010} hosted at: \url{https://zenodo.org/record/3715232\#.YLdldJNKj0p}.

This publication makes use of data products from the Wide-field Infrared Survey Explorer, which is a joint project of the University of California, Los Angeles, and the Jet Propulsion Laboratory/California Institute of Technology, and NEOWISE, which is a project of the Jet Propulsion Laboratory/California Institute of Technology. WISE and NEOWISE are funded by the National Aeronautics and Space Administration. The Pan- STARRS1 Surveys (PS1) have been made possible through con- tributions by the Institute for Astronomy, the University of Hawaii, the Pan-STARRS Project Office, the Max-Planck Society and its participating institutes, the Max Planck Institute for Astronomy, Heidelberg and theMax Planck Institute for Extraterrestrial Physics, Garching, The Johns Hopkins University, Durham University, the University of Edinburgh, the Queen’s University Belfast, the Harvard-Smithsonian Center for Astrophysics, the Las Cumbres Observatory Global Telescope Network Incorporated, the National Central University of Taiwan, the Space Telescope Science Institute, and the National Aeronautics and Space Administration under Grant No. NNX08AR22G issued through the Planetary Science Division of the NASA Science Mission Directorate, the National Science Foundation Grant No. AST-1238877, the University of Maryland, Eotvos Lorand University (ELTE), and the Los Alamos National Laboratory. Funding for SDSSIII has been provided by the Alfred P. Sloan Foundation, the Participating Institutions, the National Science Foundation, and the U.S. Department of Energy Office of Science. The SDSS-III web site is \url{http://www.sdss3.org/}. SDSSIII is managed by the Astrophysical Research Consortium for the Participating Institutions of the SDSS-III Collaboration includ- ing the University of Arizona, the Brazilian Participation Group, Brookhaven National Laboratory, Carnegie Mellon University, University of Florida, the French Participation Group, the German Participation Group, Harvard University, the Instituto de Astrofisica de Canarias, the Michigan State/Notre Dame/JINA Participation Group, Johns Hopkins University, Lawrence Berkeley National Laboratory, Max Planck Institute for Astrophysics, Max Planck Institute for Extraterrestrial Physics, New Mexico State University, New York University, Ohio State University, Pennsylvania State University, University of Portsmouth, Princeton University, the Spanish Participation Group, University of Tokyo, University of Utah, Vanderbilt University, University of Virginia, University of Washington, and Yale University.

\section*{Data Availability}
The  data  underlying  this  article  is available  from  the the LOFAR Surveys website at \url{https://lofar-surveys.org} and the code is available on GitHub at \url{https://github.com/BonnyBlu/RL-Xid/tree/main/LOFAR/DR1}.




\bibliographystyle{mnras}
\bibliography{RL-XidBib} 




\appendix
\label{sec:App}

\section{RL-Xid Results}
\label{SubSec:Results}
There are two possible outcomes of the ridgeline drawing for each source.  \textit{\revb{Completed}}, where the code has managed to find two initial directions, first points and a series of points leading from them.  \textit{\revb{Completed}} sources create individual files holding information about the ridgelines.  The other possible outcome is a \textit{\revb{Failed}}, where the code has failed at any of the possible points described above.  These \textit{\revb{Failed}} are recorded in an output file.  The sample of 991 had a 95.9 per cent success rate (41 \textit{\revb{Failed}}, see Table \ref{tab:Outcomes}); this is discussed in more detail below.

\begin{table}
    \centering
    \caption{Results from the ridgeline drawing stage of RL-Xid on the sample of 991 sources.  It gives the percentages of \textit{\revb{Completed}} and \textit{\revb{Failed}} along with the catalogue errors and analytical issues as a ratio of the whole 991 sample and of each group.}
    \begin{tabular}{p{3.5cm}|p{1.2cm}|p{1.6cm}}
    \hline
    ~ & \% Sample \textit{(991)} & \% \rev{\textit{\revb{Completed}}} \textit{(950)}  \\
    \hline
    \rev{\textit{\revb{Completed}}} \textit{(950)} & 95.9 & -  \\
    \hline
    \hline
    Masking & 1.1 & 1.2 \\
    Extended & 2.0 & 2.1 \\ 
    Looping & 5.3 & 5.5 \\
    Failed to FRII & 1.0 & 1.1 \\
    Same Direction & 0.8 & 0.8 \\    
    Not Rep & 1.8 & 1.9 \\
    Too Short & 1.7 & 1.8 \\
    \hline
    \textit{\revb{Failed}} \textit{(41)} & 4.1 & \% of \textit{\revb{Failed}} \\
    \hline
    \texttt{FloodFill} & 1.1 & 27 \\
    \texttt{Erosion} & 1.5 & 37 \\
    \textit{Initial ID Out of Region} & 0.9 & 22 \\
    \textit{Unable to find Initial Directions} & 0.5 & 12 \\
    \textit{Unable to find First Ridge Point} & 0.1 & 2 \\
    \hline
    \end{tabular}
    \label{tab:Outcomes}
\end{table}

\subsection{\textit{\revb{Completed}}}
\label{subsubsec:comp}

\begin{figure*}
    \begin{subfigure}{0.30\linewidth}
        \centering
        \includegraphics[width=\linewidth]{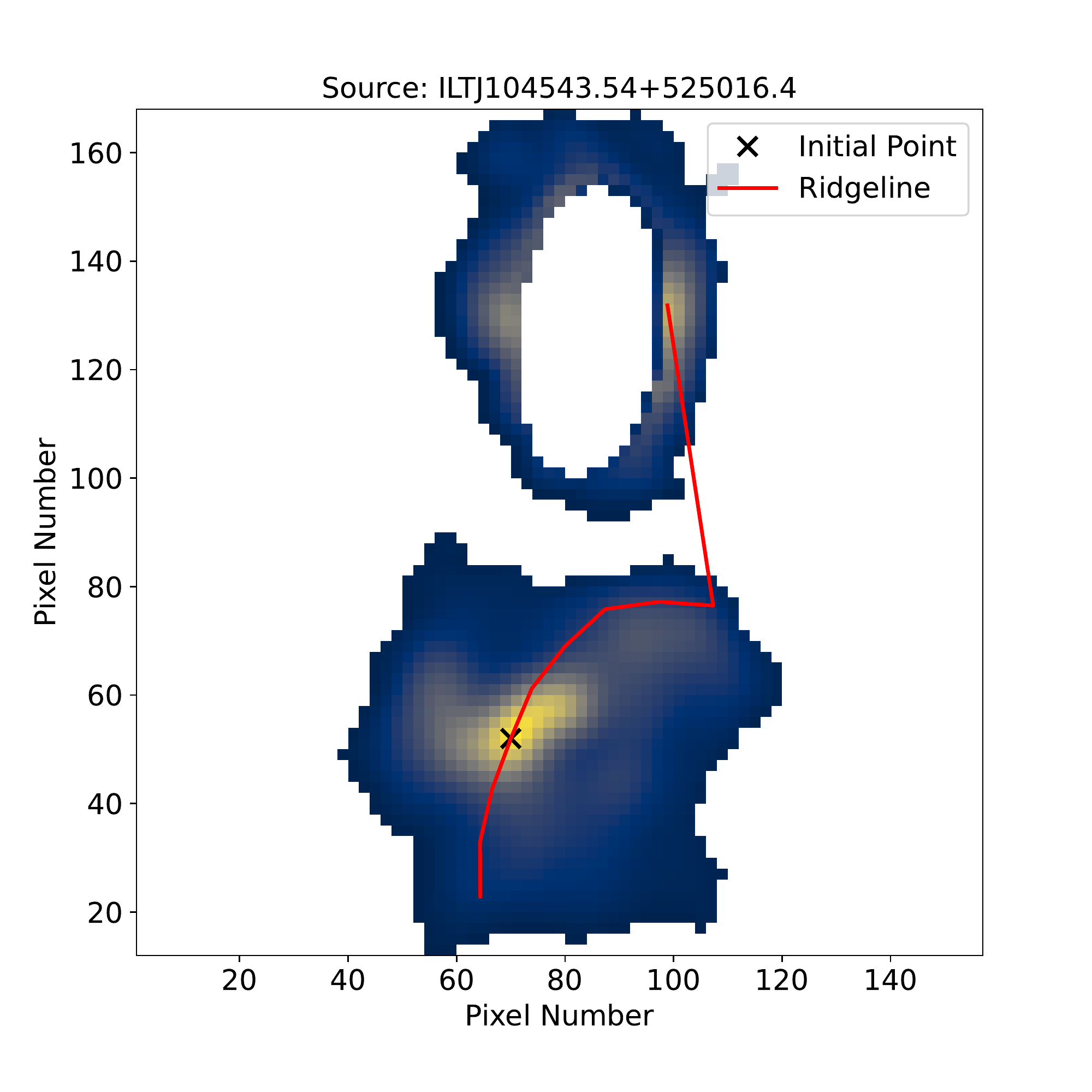}
        \caption{\label{fig:ErrExt}}
    \end{subfigure}
    \hfill
    \begin{subfigure}{0.30\linewidth}
        \centering
        \includegraphics[width=\linewidth]{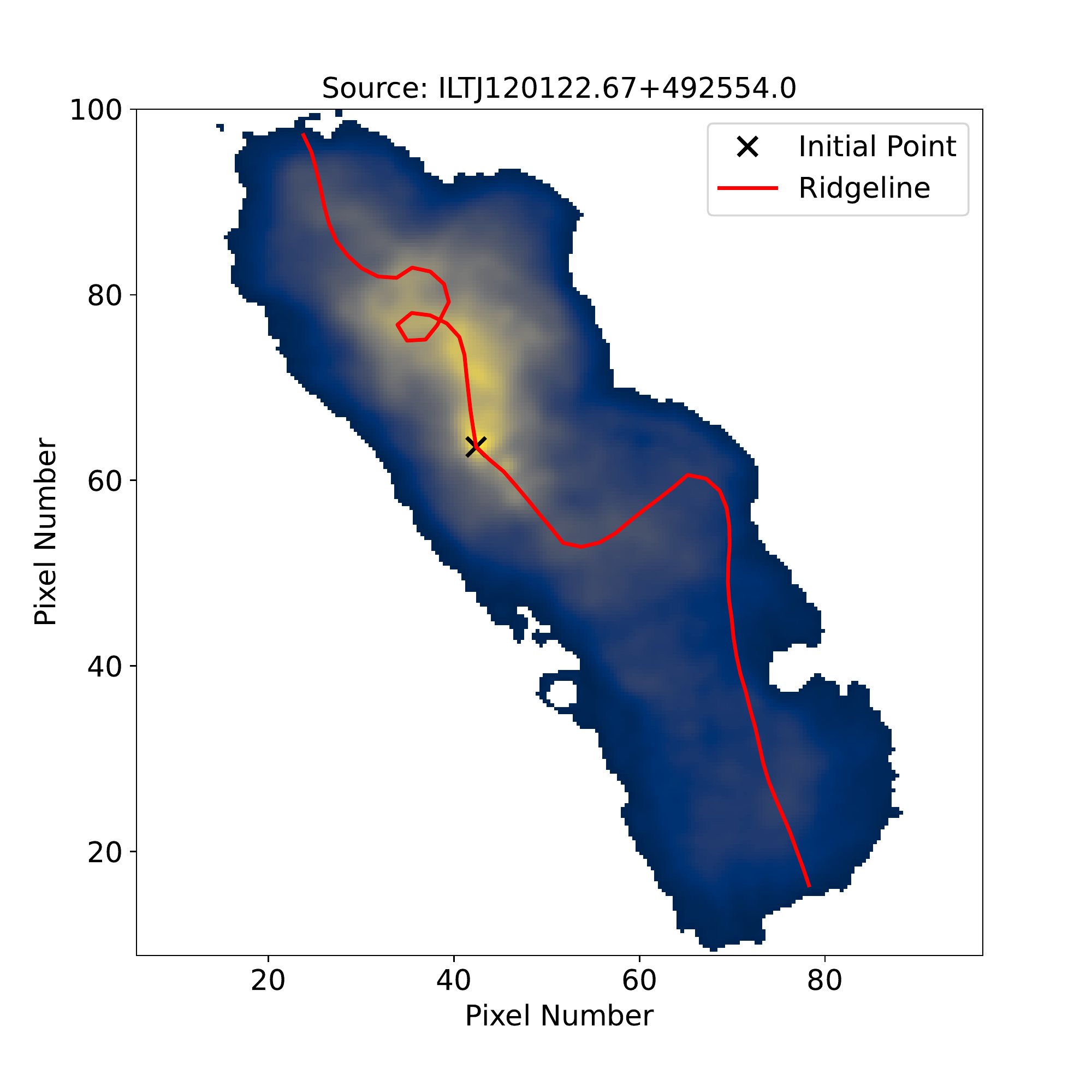}
        \caption{\label{fig:ErrFull}}
    \end{subfigure}
    \hfill
    \begin{subfigure}{0.30\linewidth}
        \centering
        \includegraphics[width=\linewidth]{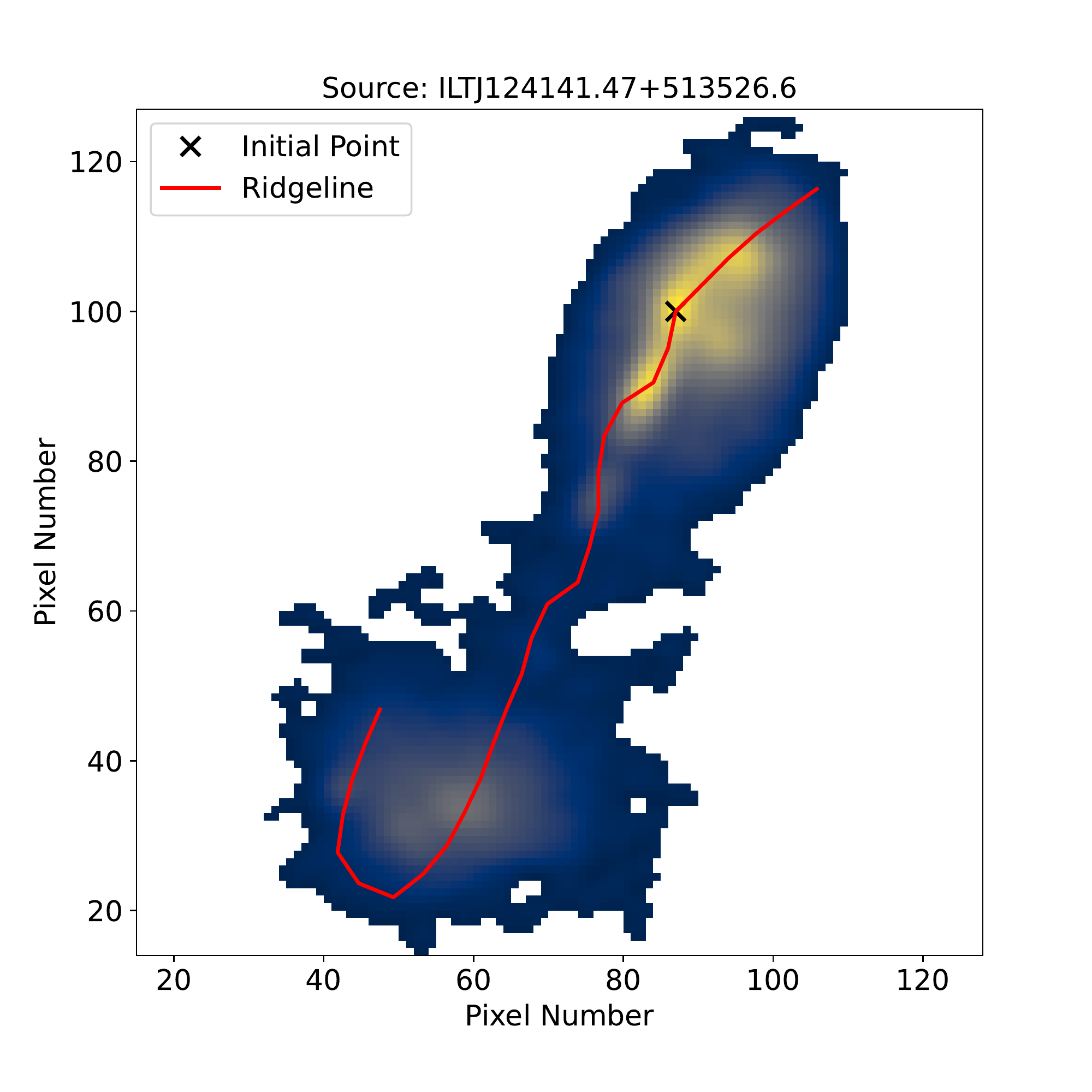}
        \caption{\label{fig:ErrEnd}}
    \end{subfigure}
    \hfill
    \begin{subfigure}{0.30\linewidth}
        \centering
        \includegraphics[width=\linewidth]{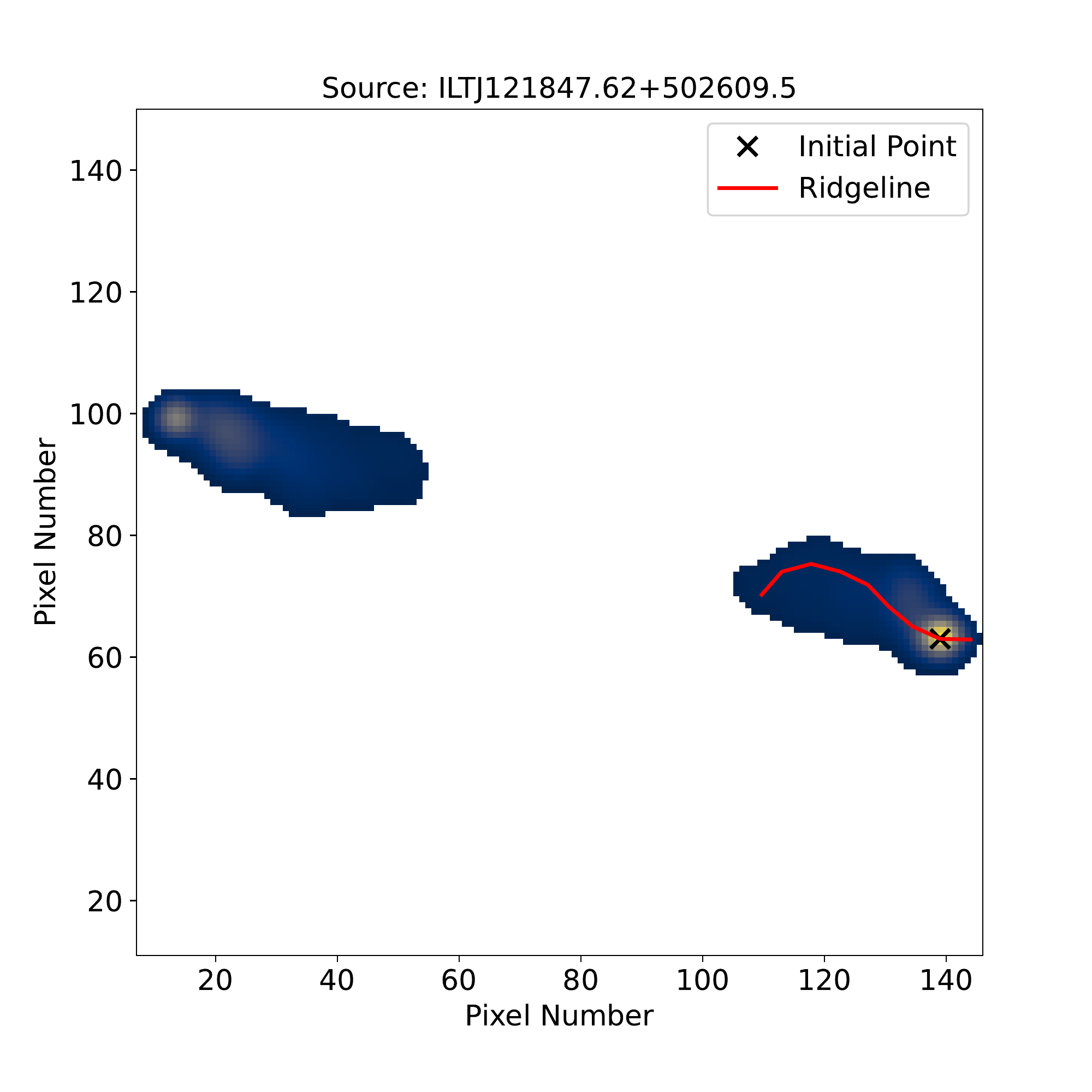}
        \caption{\label{fig:ErrFailFRII}}    
    \end{subfigure}
    \hfill
    \begin{subfigure}{0.30\linewidth}
        \centering
        \includegraphics[width=\linewidth]{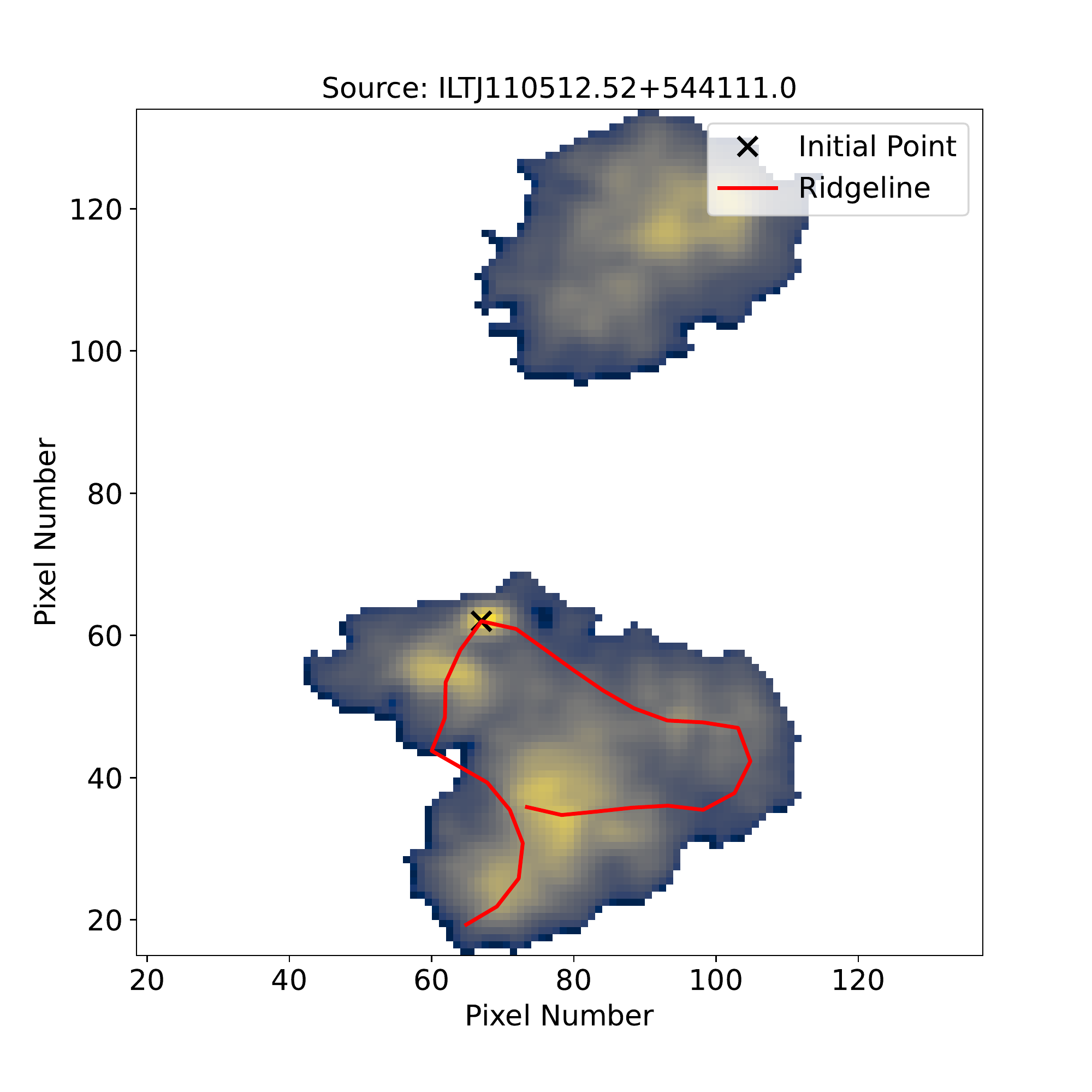}
        \caption{\label{fig:ErrBSD}}
    \end{subfigure}
    \hfill
    \begin{subfigure}{0.30\linewidth}
        \centering
        \includegraphics[width=\linewidth]{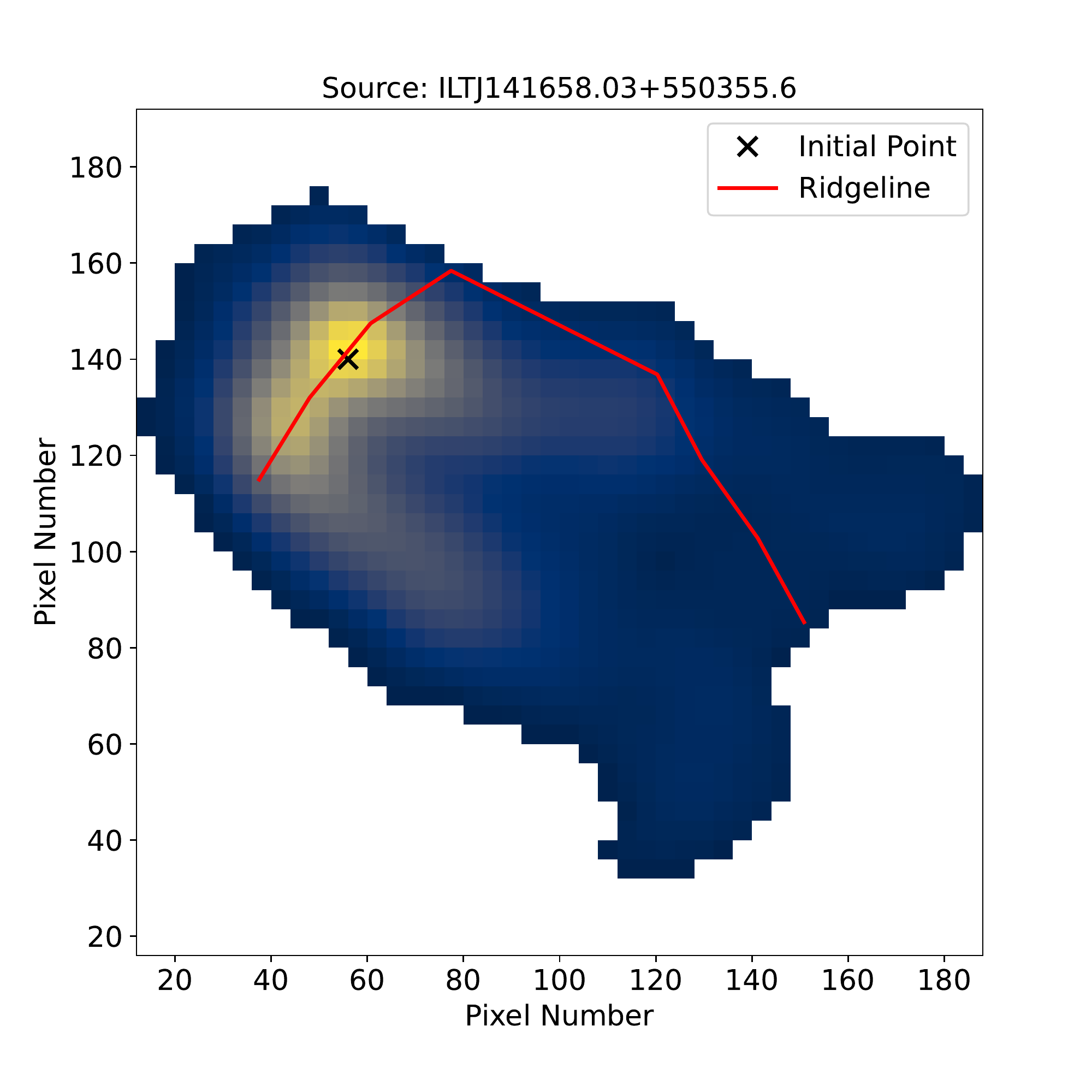}
        \caption{\label{fig:ErrNAT}}
    \end{subfigure}
    \hfill
    \begin{subfigure}{0.30\linewidth}
        \centering
        \includegraphics[width=\linewidth]{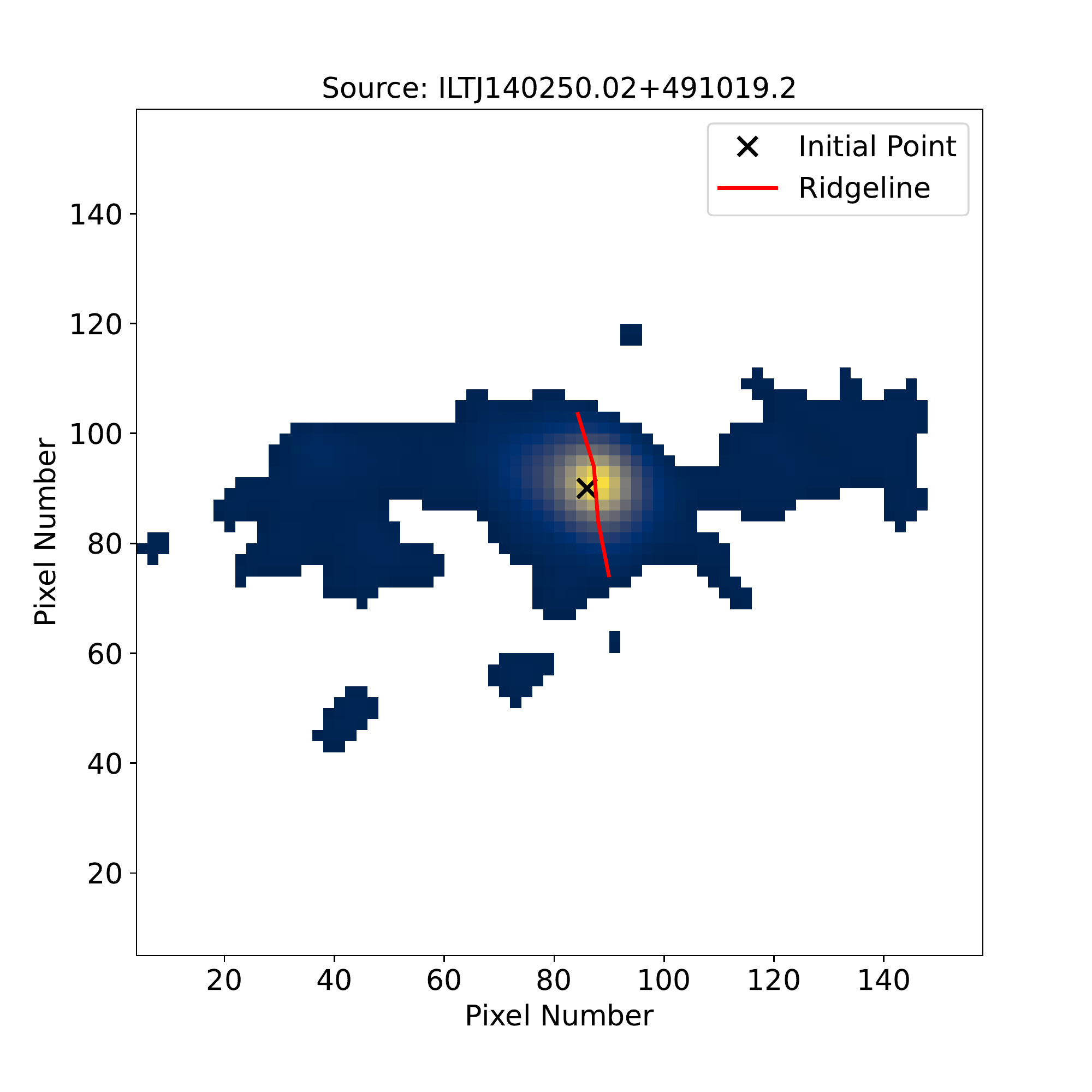}
        \caption{\label{fig:ErrNot}}
    \end{subfigure}
    \hfill
    \begin{subfigure}{0.30\linewidth}
        \centering
        \includegraphics[width=\linewidth]{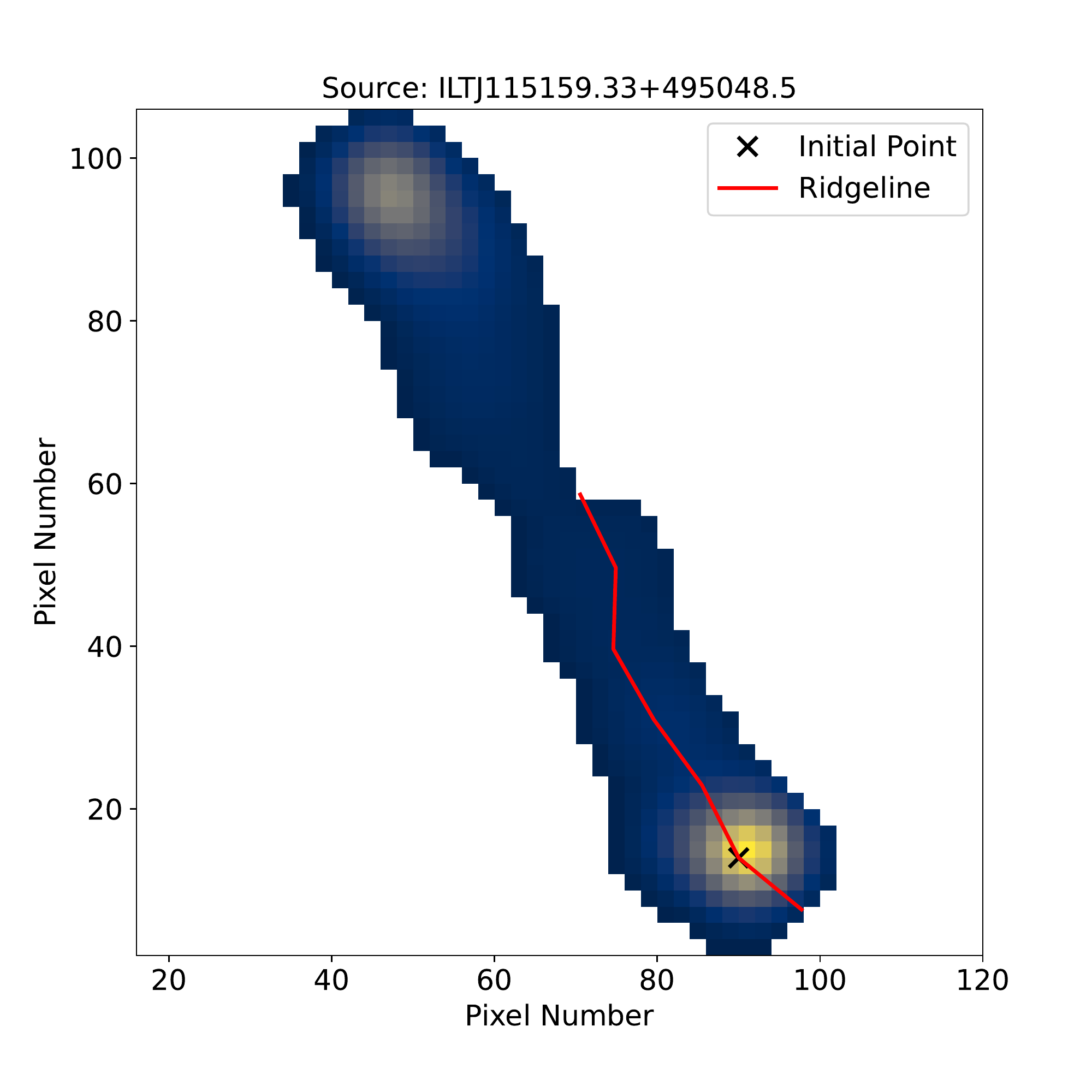}
        \caption{\label{fig:ErrShort}}
    \end{subfigure}
    \hfill
    \begin{subfigure}{0.30\linewidth}
        \centering
        \includegraphics[width=\linewidth]{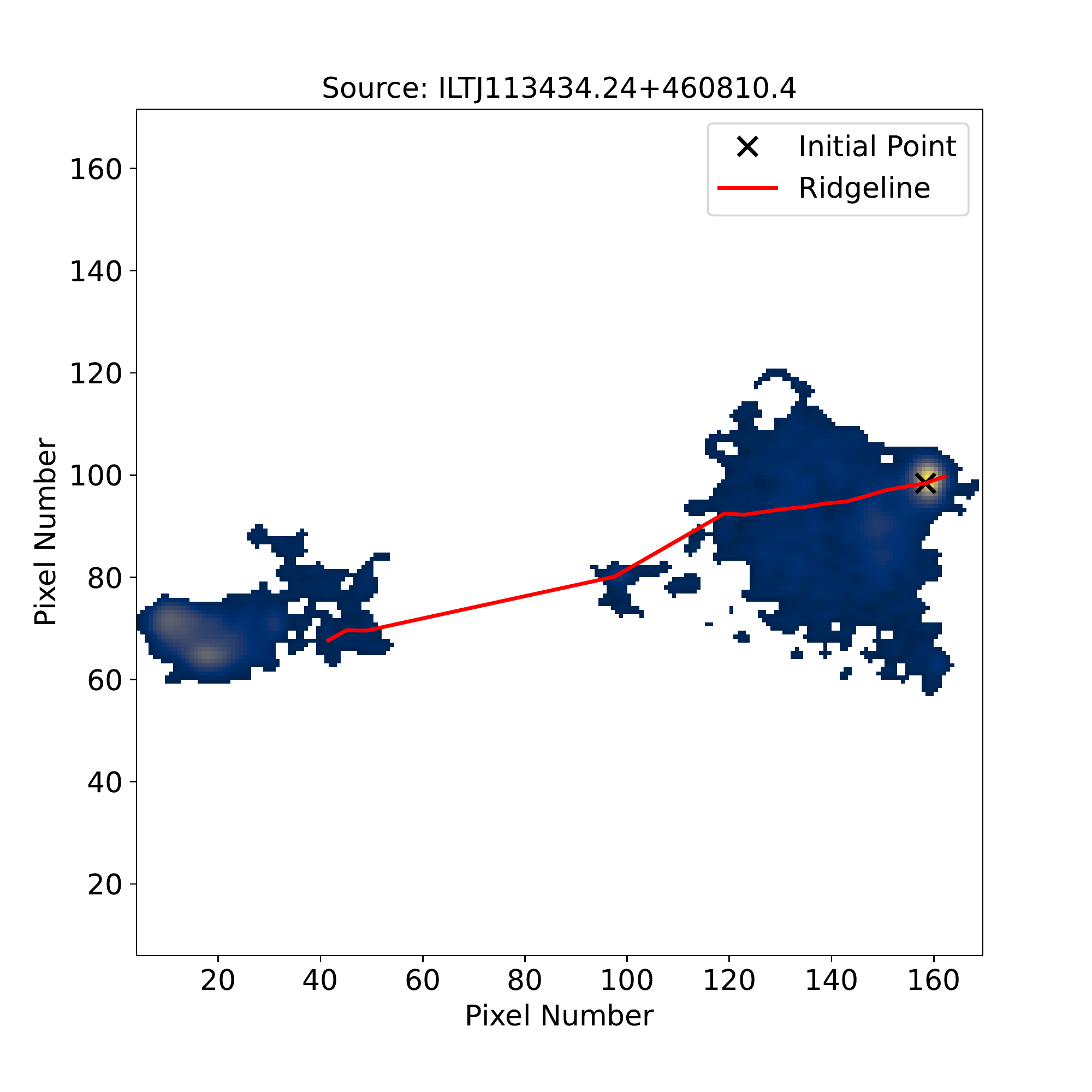}
        \caption{\label{fig:ErrShortFRII}}
    \end{subfigure}
    \hfill
    \caption{Examples of the analytical issues creating low quality ridgelines in the successful sample \revb{(1 pixel $\approx$ 1.5 arcsec)}.  These are explained in detail in Section \ref{subsubsec:comp}. \rev{(a): An example of an extended issue where the ridgeline has extended to nearby unassociated emission.  (b): A ridgeline containing a full loop.  (c): An example where the end of the ridgeline has turned back on itself for at least three steps.  (d): This ridgeline is an example where the FRII gap has not been bridged.  (e): Due to the even emission of this source the both halves of the ridgeline have travelled in the same direction.  (f): On inspection this source would appear to be a narrow angle tail source, the ridgeline drawn does not represent this.  (g): The ridgeline does not represent the source and is instead transverse to the expected direction.  (h): The ridgeline has stopped short of the end of the source.  (i): The ridgeline has bridged the gap in the emission, however it is not long enough to reach the end of the source.}}
    \label{fig:Errors}
\end{figure*}

A \textit{\revb{Completed}} source is one where the code managed to complete and \rev{draw a ridgeline} in two directions from the initial point.   A visual inspection of the \textit{\revb{Completed}} sources was carried out to check the quality of the ridgelines.  Low quality ridgelines fall into two categories; those which are caused by rare cataloguing errors and those where the code has produced a ridgeline containing analytical issues, as discussed below.  These have produced ridgelines which are faulty or misleading in some aspect.  They may not necessarily be unusable depending on the application of the ridgelines.  The catalogue errors occur with the \texttt{Erosion} and \texttt{FloodFill} functions.  Overlapping, un-associated sources or mis-identified components in the catalogue can lead to large parts of the source being masked out, see Figure \ref{fig:Masking}.  The code will have attempted to draw a ridgeline for these sources given the emission present with varying degrees of success.  These cataloguing errors are evident in 1.5 per cent of the sample; 1.1 per cent in the \textit{\revb{Completed}} sources (see Table \ref{tab:Outcomes}) and 0.4 per cent in the  \textit{\revb{Failed}}.  Examples of the analytical issues are shown in Figure \ref{fig:Errors} and described below, stating their percentage occurrence within the successfully drawn ridgelines:

\begin{figure}
    \centering
    \includegraphics[width=\columnwidth]{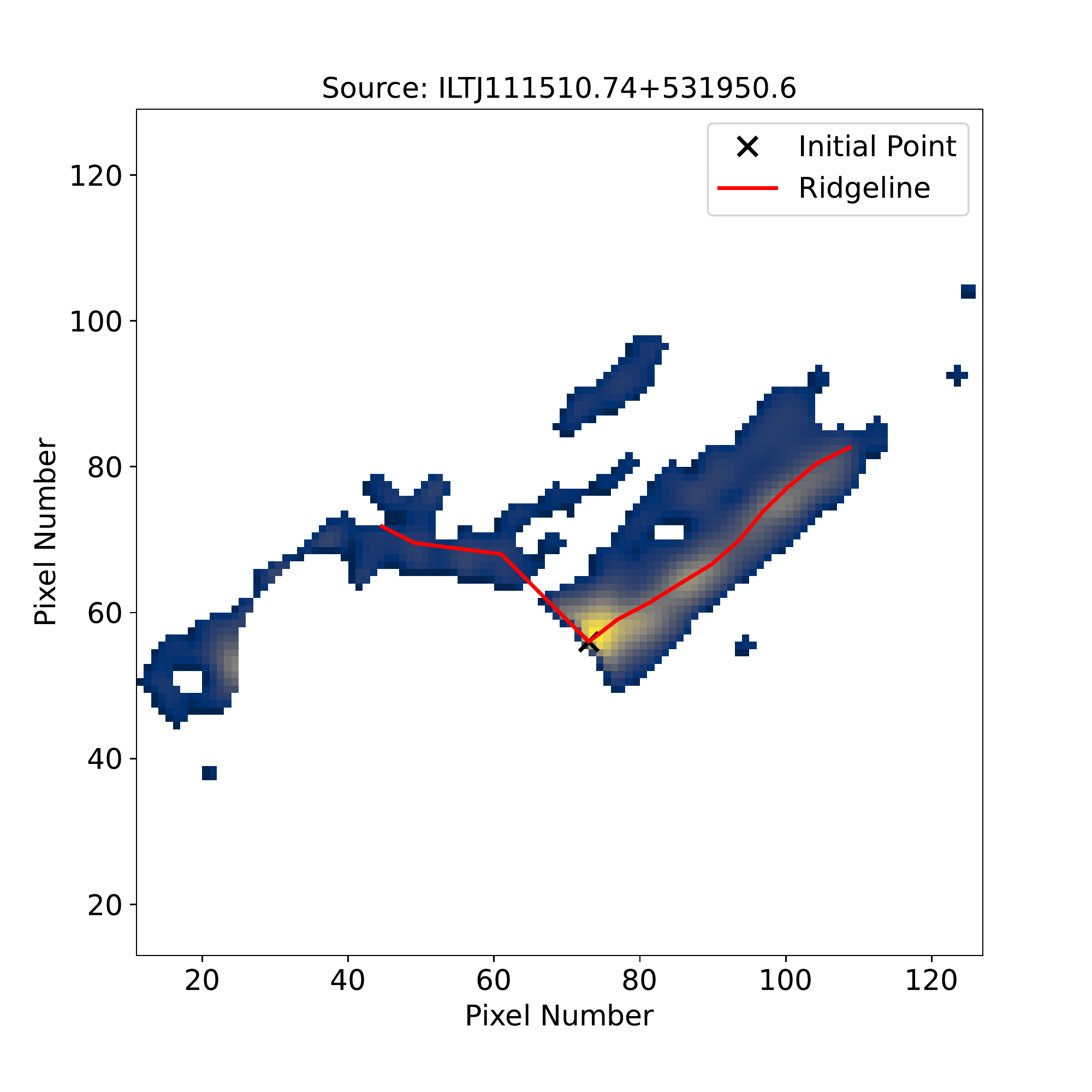}
    \caption{An example of a low quality ridgeline produced from a Masking or Flood Fill error \revb{(1 pixel $\approx$ 1.5 arcsec)}.  A large portion of the source has been masked either through an {color{red} overlapping} source or a mis-identified region, causing a completed but inaccurate attempt at a ridgeline.}
    \label{fig:Masking}
\end{figure}

\begin{description}
    \item \textit{Extended (2.1 per cent) (Figure \ref{fig:ErrExt})} – The ridgeline has jumped or extended out of the apparent source into nearby, un-associated emission.  This is due to incomplete masking or flood filling, or catalogue errors.
    \item \textit{Looping (5.5 per cent) (Figures \ref{fig:ErrFull} and \ref{fig:ErrEnd})} – The ridgeline has managed to create a full loop in its pathway or has looped back at the ends.  Full loops are more often found in large sources where the masking around the previous points fails to prevent a circular pattern from occurring.  For a source to be classed as having a loop back the ridgeline has to take three or more steps returning towards the centre, without indication in the image of an associated tail or back flow in the location of these ridge points.
    \item \textit{FRII Jump Failed (1.1 per cent) (Figure \ref{fig:ErrFailFRII})} – The ridgeline on some FRII sources did not make the jump over the central emission gap, or over a gap from the central core to one of the outer hot spots.
    \item\textit{ Both Same Direction (0.8 per cent) (Figure \ref{fig:ErrBSD})} – Both ridgelines have travelled in the same direction from the initial point. With sources that have very even emission around the initial point this can lead to both ridgelines heading in the same direction. 
    \item \textit{Not Representative (1.9 per cent) (Figures \ref{fig:ErrNAT} and \ref{fig:ErrNot})} – The ridgeline does not represent the believed pathway of the jet.  In these instances the correct path has not been determined and the ridgeline has often travelled transverse to the source as in Figure \ref{fig:ErrNot} or has been unable to correctly distinguish the separation between tails, Figure \ref{fig:ErrNAT}.
    \item \textit{Too Short (1.8 per cent) (Figures \ref{fig:ErrShort} and \ref{fig:ErrShortFRII})} – The ridgeline is too short for the source. This may occur after an FRII jump and might be because of the length restriction in place in the code (see earlier discussion in Section \ref{SubSec:TR}).
\end{description}

The breakdown of how many catalogue errors and analytical issues are present in the sample is given in Table \ref{tab:Outcomes} along with the percentages in terms of the sample as a whole and of the \textit{\revb{Completed}} sources.  From Table \ref{tab:Outcomes} the catalogue errors account for $\sim$1 per cent and analytical errors for $\sim$13 per cent of the \textit{\revb{Completed}} outcomes.  It is intended in future releases of the code to further reduce the effects of these errors.

As the initial point is the point of maximum flux \rev{density} it is expected this will coincide with the AGN core or hot spot region, and for the purposes of cross-identification may be situated near to the possible optical counterpart.  Regardless of the analytical issues the initial point is still present on the ridgeline and this allows for cross-identification to take place in the majority of cases.  All of these analytical issues are morphological in nature, therefore up to $\sim$13 per cent of the \textit{\revb{Completed}} sources could produce morphologically misleading ridgelines.  As the \textit{\revb{Completed}} ridgeline output files contain further information regarding the location of the ridge point on the array, the length of the ridgeline, and the angular change in the ridgeline, further work investing these properties could be affected in $\sim$13 per cent of the results.

\subsection{\textit{\revb{Failed}}}
\label{subsubsec:fails}
There are five possible ways for the code to produce an error; this produced 41 (4.1 per cent) instances of \textit{\revb{Failed}} sources.  Two instances, \texttt{Erosion} and \texttt{FloodFill} fail through an inability to complete their functions and are the cause of 63 per cent of the \textit{\revb{Failed}}. The remaining \textit{\revb{Failed}} are due to the ridgeline not meeting the requirements to be drawn, for example the \textit{ID Out of Region} error discussed at the end of Section \ref{SubSec:TR}.  The remaining two possibilities are simply where the source is unsuitable for the ridgeline process.  Either after multiple attempts the initial conditions are not satisfied (\textit{Unable to Find Initial Directions}) or from the initial point the first steps of the process are unable to begin (\textit{Unable to Find First Ridge Point}), see Figure \ref{fig:flow}.

As the RL-Xid algorithm lays search \revb{sectors} for two ridgelines in opposite directions from the initial point, it is possible certain morphologies might not be detected, such as narrow angle tail (NAT) sources.  Head-tail sources are narrow angle tail radio galaxies, viewed edge on, which show sharp bends very close to the core, making it hard to distinguish both tails \citep{Simon1978}.  The cores of these radio galaxies lie close to one edge of the source and may be prone to causing the code to fail due to its inability to find a first ridge point in both directions.  In order to check the effect of this, the sample was visually reviewed for head-tail sources and $\sim$ 2 per cent (22) of the sample were found to be likely head-tail candidates.  Of these the \textit{\revb{Failed}} were visually inspected for these cases and $\sim$27 per cent (6) possible candidates were found, $\sim$9 per cent (2) were classed as Erosion errors. The remaining $\sim$18 per cent (4) are possible candidates for failure as head-tail sources with an outputted error of \textit{Unable to Find the First Ridge Point}.


\bsp	
\label{lastpage}
\end{document}